\documentclass{siamltex}
\usepackage{amsmath}
\usepackage{amssymb}
\usepackage{bm}
\usepackage{bbm}
\usepackage{graphicx}
\usepackage{url}
\usepackage{xcolor}
\usepackage{cite}

\DeclareMathAlphabet{\mathantt}{OT1}{antt}{li}{it}
\DeclareMathAlphabet{\mathpzc}{OT1}{pzc}{m}{it}

\newcommand{\drt}[1]{{\color{black}#1}}

\title{Eigenvector-Based Centrality Measures for Temporal Networks
\thanks{The first two authors contributed equally to this project.
We are grateful to Mitch Keller of the Mathematics Genealogy Project \cite{mgp} for providing data. 
We thank Geoff Evans, Martin Everett, Bailey Fosdick, Echo Gao, {Sam Howison}, Florian Klimm, Flora Meng, Priya Narayan, Victor Preciado, Feng Shi, Gilbert Strang, Blair Sullivan, and Simi Wang for helpful discussions and suggestions.
We especially thank Elizabeth Leicht for many extensive discussions. DT and PJM were supported partially by the Eunice Kennedy Shriver National Institute of Child Health \& Human Development of the National Institutes of Health under Award Number R01HD075712. DT was also funded by the National Science Foundation (NSF) under Grant DMS-1127914 to the Statistical and Applied Mathematical Sciences Institute. SAM and PJM were funded by the NSF (DMS-0645369). PJM was also funded by a James S. McDonnell Foundation 21st Century Science Initiative - Complex Systems Scholar Award (\#220020315). MAP was supported by the FET-Proactive project PLEXMATH (\#317614) funded by the European Commission, a James S. McDonnell Foundation 21st Century Science Initiative - Complex Systems Scholar Award (\#220020177), and the EPSRC (EP/J001759/1). AC was funded by the NSF under Grant IIS-1452718. Any content is solely the responsibility of the authors and do not necessarily reflect the views of any of the funding agencies.
}}

\author{Dane Taylor\thanks{Carolina Center for Interdisciplinary Applied Mathematics, Department of Mathematics, University of North Carolina, Chapel Hill, NC 27599-3250, USA; and Statistical and Applied Mathematical Sciences Institute (SAMSI), Research Triangle Park, NC, 27709, USA}
\and Sean A. Myers\thanks{Carolina Center for Interdisciplinary Applied Mathematics, Department of Mathematics, University of North Carolina, Chapel Hill, NC 27599-3250, USA (Current address: Department of Economics, Stanford University, Stanford, CA 94305-6072, USA)}
\and Aaron Clauset\thanks{Department of Computer Science, University of Colorado, Boulder, CO 80309, USA; Santa Fe Institute, Santa Fe, NM 87501, USA; and BioFrontiers Institute, University of Colorado, Boulder, CO 80303, USA}
\and Mason A. Porter\thanks{Mathematical Institute, University of Oxford, OX2 6GG, UK; CABDyN Complexity Centre, University of Oxford, Oxford OX1 1HP, UK; and Department of Mathematics, University of California, Los Angeles, CA 90095, USA}
\and Peter J. Mucha\thanks{Carolina Center for Interdisciplinary Applied Mathematics, Department of Mathematics, University of North Carolina, Chapel Hill, NC 27599-3250, USA}
}

\begin{document}
\maketitle 

\begin{abstract}
Numerous centrality measures have been developed to quantify the importances of nodes in \drt{time-independent} networks, and many of them can be expressed as the leading eigenvector of some matrix. With the increasing availability of network data that changes in time, it is important to extend \drt{such} eigenvector-based centrality measures to time-dependent networks. In this paper, we introduce a principled generalization \drt{of network centrality measures} that is valid for any eigenvector-based centrality. We consider a temporal network with $N$ nodes as a sequence of $T$ layers that describe the network during different time windows, and we couple centrality matrices for the layers into a \emph{supra-centrality} matrix of size $NT\times NT$ whose dominant eigenvector gives the centrality of each node $i$ at each time $t$. We refer to this eigenvector and its components as a \emph{joint centrality}, as it reflects the importances of both the node $i$ and the time layer $t$. We also introduce the concepts of {\emph{marginal}} and {\emph{conditional}} centralities, which facilitate the study of centrality trajectories over time. We find that the strength of coupling between layers is important for determining multiscale properties of centrality, such as localization phenomena and the time scale of centrality changes. In the \drt{strong-coupling regime}, we derive expressions for {\emph{time-averaged centralities}}, which are given by the zeroth-order terms of a singular perturbation expansion. We also study first-order terms to obtain \emph{first-order-mover scores}, which concisely describe the magnitude of nodes' centrality changes \drt{over time}. As examples, we apply our method to three empirical temporal networks: the United States Ph.D. exchange in mathematics, costarring relationships among top-billed actors during the Golden Age of Hollywood, and citations of decisions from the United States Supreme Court. 
\end{abstract}

\begin{keywords}
Temporal networks, Eigenvector centrality, Hubs and authorities, Singular perturbation, Multilayer networks, Ranking systems
\end{keywords}

\begin{AMS} 91D30, 05C81, 94C15, 05C82, 15A18 \end{AMS}

\pagestyle{myheadings}
\thispagestyle{plain}
\markboth{D. TAYLOR {\it et al.}}{Eigenvector-Based Centrality for Temporal Networks}

\section{Introduction}\label{sec:intro}

\drt{
The study of centrality \cite{newman2010}---that is, determining the importances of different nodes, edges and other structures in a network---has widespread applications in the identification and ranking of important agents (or interactions) in a network. These applications include ranking sports teams or individual athletes \cite{monthly,saavedra2010,chartier2013}, the identification of influential people \cite{kempe2003}, critical infrastructures that are susceptible to congestion\cite{holme2003congestion,guimera2005worldwide}, impactful United States Supreme Court cases \cite{leicht-citation2007,fowler2007b,fowler2008}, genetic and protein targets \cite{jeong2001lethality}, impactful scientific journals \cite{bergstrom2008eigenfactor}, and much more.  An especially important type of centrality measure are ones that arise as a solution of an eigenvalue problem \cite{gleich2014}, with nodes' importances given by the entries of the dominant eigenvector of a so-called \emph{centrality matrix}. Prominent examples include eigenvector centrality \cite{bonacich1972}, PageRank \cite{Brin1998conf,pagerank} (which provides the mathematical foundation to the Web search engine Google \cite{langville2006,gleich2014}), hub and authority (i.e., HITS) scores \cite{kleinberg1999}, and Eigenfactor \cite{bergstrom2008eigenfactor}. However, despite the fact that real-world networks change with time \cite{holme2011,holme2013,holme2015}, most methods for centrality (and node rankings that are derived from them) have been restricted to time-independent networks. Extending such ideas to time-dependent networks (i.e., so-called \emph{temporal networks}) has recently become a very active research area
\cite{tang2010,alsayed2015,pan2011,kim2012b,williams2015,lerman2010centrality,motegi2012,grindrod2011communicability,estrada2013,Grindrod_Higham_2014,taro2015,rocha2014,rossi2012,you2015distributed}.}

\begin{figure}[t!]
\centering
\includegraphics[width=.95\linewidth]{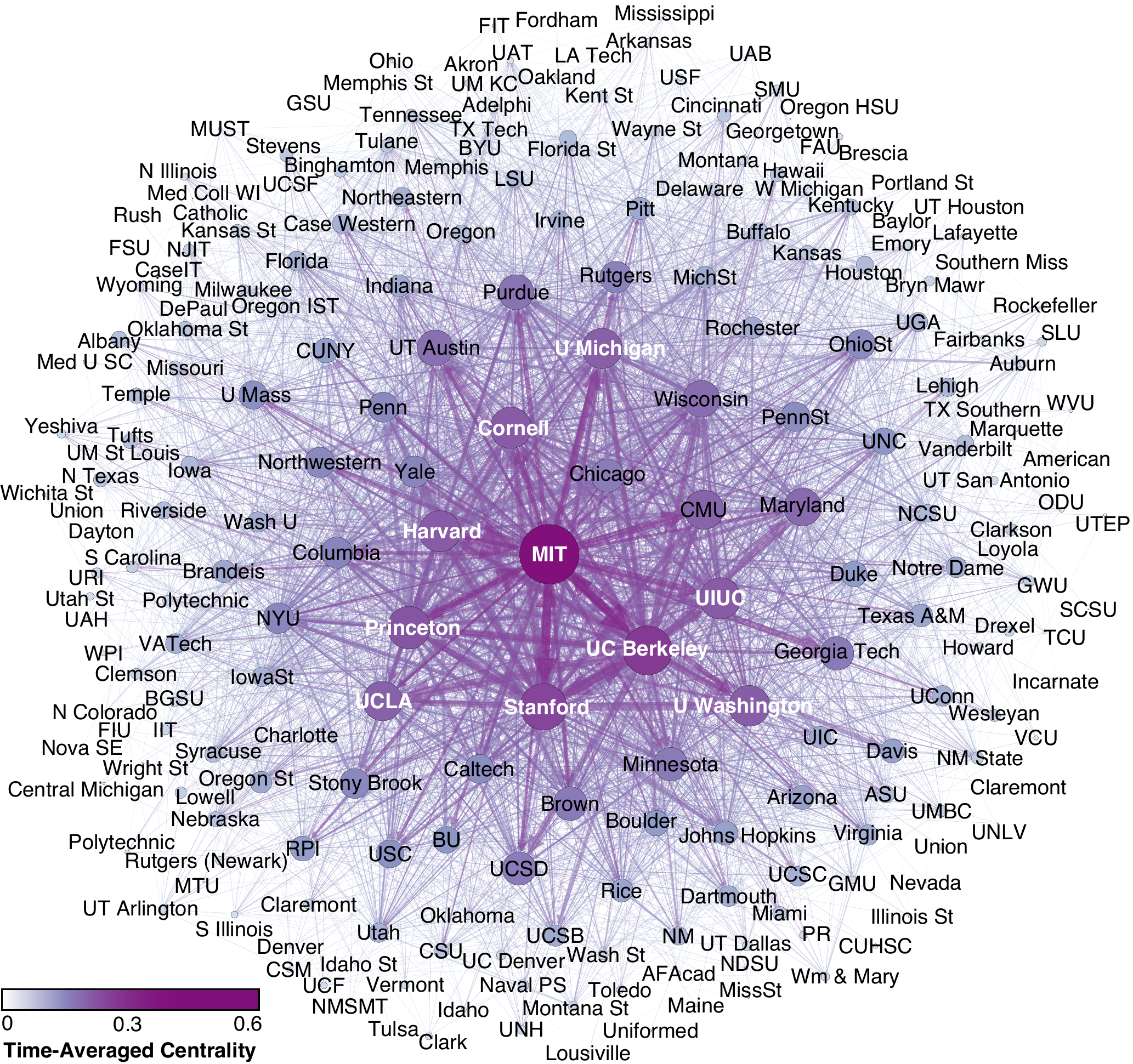}
\caption{
As one case study for our generalization of eigenvector-based centralities to temporal networks, we examine a multilayer temporal network that we construct from data obtained from the Mathematics Genealogy Project (MGP) \cite{mgp}. \drt{(See the Supplementary Material or \cite{MGP_data} to download the temporal network data.)} We study a temporal network with $T=65$ time layers (corresponding to the years 1946--2010), in which a given edge $j\to i$ in time layer $t$ signifies the number of graduating doctoral students in year $t$ at university $i$ who later \drt{supervise} a graduating doctoral student at university $j$. The nodes' sizes and colors indicate what we call ``time-averaged centrality'' (see Sec.~\ref{sec:TimeAve}), which we calculate for the type of centrality matrix known as an authority matrix \cite{kleinberg1999}. We do not show self-edges. Because visualizing a temporal network is difficult \drt{\cite{de2014muxviz}}, we show the network that we obtain by aggregating the adjacency matrices across time layers. Studying such an aggregated network neglects the time-ordered structure \drt{of} a temporal network. In our paper, we study trajectories of node centralities (which represent importances) over time. See Sec.~\ref{sec:mgp} for additional discussion of the MGP network. \drt{(We created this image using the software Gephi \cite{ICWSM09154}.)}
}
\label{fig:MGP_network}
\end{figure}

In the present paper, we develop a generalization of eigenvector-based centralities to ordered multilayer networks such as temporal networks. Akin to multilayer modularity \cite{mucha2010}, a critical underpinning of our approach is that we treat inter-layer and intra-layer connections as fundamentally different \cite{kivela2014,de2013b}. For the purpose of generalizing centrality, we implement this idea by coupling centrality matrices for temporal layers in a {\it supra-centrality matrix}. For a temporal network with $N$ nodes and $T$ time layers, we obtain a supra-centrality matrix of size $NT\times NT$. To use this matrix, we represent the temporal network as a sequence of network layers, which is appropriate for discrete-time temporal networks and continuous-time networks whose edges have been binned to produce a discrete-time temporal network. \drt{Importantly, our {methodology} is independent of the particular choice of centrality matrix, so it can be used, for example, with ordinary eigenvector centrality \cite{bonacich1972}, hub and authority scores \cite{kleinberg1999}, PageRank \cite{pagerank}, or any other centralities that are given by components of the dominant eigenvector of a matrix.} 

The dominant eigenvector of a supra-centrality matrix characterizes the \emph{joint centrality} of each node-layer pair $(i,t)$---that is, the centrality of node $i$ at time step $t$---and \drt{thus} reflects the importances of both node $i$ and layer $t$. We also introduce the concepts of \emph{marginal centrality} and \emph{conditional centrality}, which allow one to (1) study the decoupled centrality of just the nodes (or just the time layers) and (2) study a node's centrality with respect to other nodes' centralities at a particular time $t$ (i.e., the centralities are conditional on a particular time layer). These notions make it possible to develop a broad description for studying nodes' centrality trajectories across time. Although we develop this formalism for temporal networks, we note that our approach is also applicable to multiplex and general multilayer networks, which are two additional scenarios in which the generalization of centrality measures is important. (See, e.g., \cite{Halu_Mondragon_Panzarasa_Bianconi_2013,sola2013eigenvector,sole2014centrality} for multiplex centralities and \cite{manlio2015-centrality,sole2015} for general multilayer centralities.) 

Similar to the construction of \drt{supra-adjacency matrices} \cite{mucha2010,kivela2014,supra2013}
\drt{(see Sec.~\ref{sec:naive} for definition)}, 
we couple nearest-neighbor temporal layers using inter-layer edges of weight $\omega$, which leads to a family of centrality measures that are parameterized by $\omega$. The parameter $\omega$ controls the coupling of each node's eigenvector-based centrality through time and can be used to tune the extent to which a node's centrality changes over time. The limiting cases $\omega\to0^+$ and $\omega\to\infty$ are particularly interesting, as the former represents the regime of complete decoupling of the layers and the latter represents a regime of dominating coupling of the layers. As part of the present paper, we conduct a perturbative analysis for the $\omega\to\infty$ \drt{limit. (For notational convenience, we study $\epsilon\to0^+$ for $\epsilon=1/\omega$.)} This allows us to derive principled expressions for the \drt{nodes' \emph{time-averaged centralities}} (given by the zeroth-order expansion) and \emph{first-order-mover scores}, which are derived from the first-order expansion. Time-averaged centrality ranks nodes so that their centralities are constant across time, and first-order-mover scores rank nodes according to the extent to which their centralities change in time. The computation of both time-averaged centralities and first-order-mover scores  \drt{can be} very efficient, because they only require the numerical solution of linear-algebraic problems of size $N \times N$, which is ordinarily much smaller than the full supra-centrality matrix of size $NT \times NT$. Moreover, our perturbative approach also alleviates the need to demand a particular choice of intra-layer coupling weight $\omega$.
\drt{Note that our methodology is a retrospective analysis and complements alternative temporal generalizations that take a causal approach.}

\drt{The remainder of our paper is organized as follows. 
In Sec.~\ref{sec:back_and_naive}, we provide further background information.
In Sec.~\ref{sec:method}, we present our generalization of eigenvector-based centralities for temporal networks.} 
In Sec.~\ref{sec:TimeAve}, we derive principled expressions for time-averaged \drt{centralities} and first-order-mover scores based on a singular perturbation expansion. 
In Sec.~\ref{sec:num}, we illustrate our approach using three examples from empirical data: \drt{Ph.D. exchange using data from} the Mathematics Genealogy Project \cite{mgp} (see Fig.~\ref{fig:MGP_network}), top billing in the Golden Age of Hollywood using the Internet Movie Database (IMDb) \cite{IMDb}, and citations of United States Supreme Court decisions \drt{\cite{SCD,fowler_url}.} We conclude in Sec.~\ref{sec:conclusion} and provide further details of our perturbation expansion in \drt{the} appendix.

\section{Background Information and a Naive Approach}\label{sec:back_and_naive}

\drt{In Sec.~\ref{sec:background}, we provide additional background information. In Sec.~\ref{sec:naive}, we discuss a naive approach for temporal eigenvector centrality that motivates our  approach, and introduces our mathematical notation and terminology.}

\subsection{\drt{Background Information}\label{sec:background}}

The analysis of large networks is ubiquitous in science, engineering, medicine, and numerous other areas \cite{newman2010}. In the social sciences, for example, the abundance of data that describes the social behavior of individuals in academia \cite{mgp,caplow2001,myers2011,paneretos2012,burris2004,clauset2015}, show business \cite{sarkar2014}, politics \cite{leicht-citation2007,fowler2007b,fowler2008,bommarito2010,congshort}, and just about every other arena offers exciting avenues for the quantitative study of social systems. For these and many other applications, it is important to develop (and improve) mathematical techniques to extract concise and intuitive information from large network data. From the interdisciplinary pursuit of what is now often called {\emph{network science,}} we know---from theory, computation, and data analysis---that many network properties (e.g., degree heterogeneity, local clustering, community structure, and others \cite{newman2010}) have significant effects on dynamical processes on networks (e.g., information dissemination and disease spreading) \cite{porter2014,wassermanfaust,colizza2006}. Although the vast majority of research in network science has focused on \drt{time-independent} networks, increased effort in recent years has aimed to generalize network analyses to ``temporal networks'' \cite{holme2011,holme2013,holme2015,vlado2014,kempe2002}, in which network entities and/or interactions change in time.

In the study of \drt{time-independent} networks, numerous centrality measures have been developed to try to quantify the relative importances of nodes \cite{wassermanfaust,newman2010}. There are seemingly as many centrality measures as applications \cite{katz,betweenness,bonacich1972,bonacich,estrada2008,benzi2013}, and different types of centrality are appropriate for different situations. Importantly, one can construct centrality measures not only from the direct consideration of network structure but also based on studying an appropriate dynamical process on a network \cite{pagerank,restrepo2006,ghosh2014}.
\drt{In our paper,} we focus on eigenvector-based centralities. Although eigenvectors can obviously be used in different ways to introduce different notions of centrality, \drt{we reserve the term \emph{eigenvector-based centrality}} to refer only to centrality measures in which the nodes' centralities are given by the entries of the dominant eigenvector of a matrix, which \drt{we call} a \emph{centrality matrix}.  Centrality matrices include a network's adjacency matrix ${\bf A}$ (which indicates which nodes share a common edge) and various functions of ${\bf A}$ \cite{benzi2013,estrada2010}, such as the hub and authority matrices \cite{kleinberg1999} and PageRank matrices \cite{pagerank,langville2006,gleich2014}. Eigenvector-based centralities reflect a network's global structure and are often preferable to other types of centralities, because there are a large number of computationally efficient algorithms for computing spectral properties of matrices (e.g., the power method for computing the dominant eigenvector \cite{golub}). An eigenvector-based centrality is also a key component to the function of major Web search engines, including Google \cite{pagerank,kleinberg1999,gleich2014}.

Although the study of centrality measures \drt{in time-independent} networks has been insightful for numerous applications, most networks are time-dependent, and it is important to generalize centrality measures for temporal networks. This has been a very active area of research in recent years. Many approaches consider a temporal network as a sequence of network layers, and initial studies examined centralities of such uncoupled individual network layers \cite{braha2006}. It was found, for example, that in the absence of coupling between network layers, centrality scores can fluctuate significantly from one time step to the next due to stochasticity in the appearance and disappearance of edges. 
\drt{Rankings can of course change over time (either slowly or rapidly). To give an example that is directly relevant to one of our case studies, it was demonstrated recently that uncoupled university rankings can fluctuate significantly from year to year \cite{sorz2015}, which we consider to be a suboptimal description for the temporal evolution of departmental prestige.}

The most popular \drt{temporal extensions of centrality measures} involve considering notions of centrality based on paths in a \drt{time-independent} network and \drt{generalizing} them using \emph{time-respecting paths} in a temporal network \cite{kossinets2008structure,kostakos2009temporal}. 
\drt{We point out temporal extensions to PageRank that have been introduced to counteract age bias \cite{walker2007ranking,Mariani2016,you2015distributed}, capture temporal changes for the external interests of nodes via dynamic teleportation \cite{rossi2012}, and study random walks taking place on temporal networks \cite{rossi2012}.}
\drt{We also highlight temporally extended versions of} betweenness centrality \cite{tang2010,kim2012b,williams2015,alsayed2015}, closeness centrality \cite{tang2010,pan2011,kim2012b,williams2015}, Bonacich centrality \cite{lerman2010centrality}, win/lose centrality \cite{motegi2012}, communicability \cite{grindrod2011communicability,estrada2013,Grindrod_Higham_2013}, Katz centrality \cite{Grindrod_Higham_2014}, and coverage centrality \cite{taro2015}. 
These research efforts have extended centrality measures in different ways, \drt{which one can relate partly to the fact that one can define time-respecting paths in different ways. A} time-respecting path can allow one, several, or an unlimited number of edge traversals during a particular time step. Moreover, the length of a time-respecting path---which one can use to provide a notion of ``distance'' between nodes---can also be measured in different ways \cite{williams2015}. \drt{For instance,} the length of a time-respecting path can describe the number of edges that are traversed by the path, latency between the initialization and termination times of the path, or \drt{a} combination of such ideas. For temporal networks, \drt{even the} notion of a ``shortest path'' lacks an unambiguous definition, \drt{which subsequently can also introduce ambiguity into dynamical processes and network measures derived from them.}

\drt{One common feature of existing generalizations of centralities} for temporal networks is that they illustrate the importance of studying an entire temporal \drt{network, as opposed to aggregating} temporal layers into a single (time-independent) network or \drt{analyzing} the time layers in isolation from one another \cite{holme2011,holme2015}. 
Specifically, studying a layer-aggregated network prevents one from studying \emph{centrality trajectories} (i.e., how centrality changes \drt{over} time), and studying the time layers in isolation does not account for the temporal orderings of edges, which can be crucial for determining centralities in a temporal setting \cite{Grindrod_Higham_2013,Grindrod_Higham_2014,estrada2013,mariani2015,fenu2015}. 
To provide additional context, we highlight that dynamical processes can behave vastly differently on temporal versus layer-aggregated networks.\footnote{As discussed in, e.g., \cite{kivela2014,de2013b} and several references therein---and more recently in \cite{deford2015}---a similar issue arises more generally in multilayer networks, and one must also take into account the effects of inter-layer edges (which are fundamentally different from \drt{intra-layer edges) when defining a dynamical process on a multilayer network.}}
For example, a random walk---a process on which many eigenvector-based centralities rely---on a temporal network is affected fundamentally by the temporal ordering and time scale of the appearances and disappearances of edges 
\cite{holme2011,holme2013,holme2015,hoffmann2012,hoffmann2013,jo-prx2014}. 
Rankings, such as eigenvector-based centralities, that are derived from such dynamics are, in turn, affected fundamentally by the temporal structure of the networks, and aggregation (as well as isolation) can lead to misleading or even simply wrong results. Additionally, if one starts with a Markovian process on a temporal network and then aggregates the network, then in general one does not obtain a Markovian process \cite{holme2013}, so fundamental (and often desirable) properties of a dynamical process can be destroyed as a byproduct of neglecting a network's inherent temporal structure.  

\drt{Finally, although extending centrality to temporal networks has become a very active research area, one should expect different generalizations of eigenvector-based centralities to be appropriate for different applications. Existing papers have not always been clear about the modeling assumptions and tradeoffs of their approaches.}

\begin{table}[b!]
\caption{A summary of our mathematical notation.}
\centering 
\begin{tabular}{c c c} 
\hline\hline 
Typeface & Class & Dimensionality  \\ [0.5ex] 
\hline 
$\mathbb{M}$ & matrix & $NT\times NT$   \\ 
$\mathbf{M}$ & matrix & $N\times N$  \\
$\bm{M}$ & matrix & $T\times T$   \\
$\mathbbm{v}$ & vector & $NT\times 1$  \\
$\mathbf{v}$ & vector & $N\times 1$   \\ 
$\bm{v}$ & vector & $T\times 1$   \\ 
$M_{ij}$ & scalar & 1   \\ 
$v_i$ & scalar & 1   \\
[1ex] 
\hline 
\end{tabular}\\
\label{table:notation} 
\end{table}

\subsection{Naive Generalization of Eigenvector Centrality for Temporal Networks}\label{sec:naive}

\drt{Before we present our primary approach in Sec.~\ref{sec:method}, it is instructive to consider one possible way to generalize eigenvector-based centralities. This example motivates our approach and introduces mathematical notation and terminology. Importantly, it is naive in that it does not treat intra-layer edges and inter-layer edges as distinct types of edges, which causes problems when the network layers are strongly coupled.}


\drt{We use a multilayer representation of networks and} seek to identify the most central nodes of a temporal network with $N$ distinct nodes {(i.e., vertices or actors)} across $T$ time layers. 
We specify the network edges with a \drt{node-by-node-by-time} ($N\times N\times T$) adjacency tensor in which nonzero elements $A_{ij}^{(t)}$ indicate the presence and weight of the edge from node $i$ to node $j$ in time layer $t$. \drt{That is}, the adjacency matrix at time $t$ is given by $\mathbf{A}^{(t)}$. See Table \ref{table:notation} for a summary of our mathematical notation. We refer to node $i$ in layer $t$ as a \emph{node-layer pair} $(i,t)$ and node $i$ (regardless of layer) as a \emph{physical node}. We are \drt{interested particularly} in understanding the physical nodes' \emph{centrality trajectories} through time.

It is tempting to reshape a network's associated adjacency tensor into an $NT\times NT$ \emph{supra-adjacency matrix} 
\begin{equation}
	{\mathbb{A}} = \left[ \begin{array}{cccc} 
 \mathbf{A}^{(1)} & \omega\mathbf{I} & 0 & \cdots\\ 
 \omega\mathbf{I} & \mathbf{A}^{(2)} & \omega\mathbf{I} &\ddots\\ 
 0 & \omega\mathbf{I} & \mathbf{A}^{(3)} &\ddots\\
 \vdots & \ddots &\ddots&\ddots\\
 \end{array}
 \right]\,,
 \label{eq:supra}
\end{equation}
which represents a collection of both the temporal network edges (i.e., intra-layer edges) \drt{and} the ``{identity edges}'' (which are inter-layer edges) that couple the node-layer pairs $\{(i,t)\}$ for the same physical node $i$ across the $T$ network layers. The identity edges of weight $\omega$ attempt to weight the persistence of a physical node through time by enforcing an identification with itself at consecutive times \cite{bazzi2015}. 
\drt{When there are inter-layer edges only between different instances of the same physical node, a multilayer network is said to exhibit  {diagonal coupling}, and the use of a constant $\omega$ across all such inter-layer edges is sometimes known as  {layer coupling} \cite{kivela2014}.} We restrict our attention to nonnegative inter-layer coupling $\omega\geq 0$. (One can consider $\omega<0$ to drive negative coupling between layers, but we do not examine such values in our applications.) One can construe $\omega$ as a parameter to tune interactions between network layers \cite{muchaporter2010,bassett2013,bazzi2015}. In the limit $\omega\to0^+$, the layers become uncoupled; in the limit $\omega\to\infty$, the layers are so strongly coupled that inter-layer weights dominate the intra-layer connections.

\drt{We also restrict ourselves to nearest-neighbor coupling of temporal layers, as we place the identity inter-layer edges only between node-layer pairs, $(i,t)$ and $(i,t\pm1)$, that are adjacent in time (where the $t = 0$ and $t = T$ layers have inter-layer edges to one other layer rather than two). This results} in the block structure in Eq.~\eqref{eq:supra}.
Equivalently, we write
\begin{equation}
	{\mathbb{A}}  = \diag\left[\mathbf{A}^{(1)},\dots, \mathbf{A}^{(T)}\right] + \bm{A}^{(\mathrm{chain})} \otimes \omega\mathbf{I}\,,\label{eq:sup_def}
\end{equation}
where $\otimes$ denotes the Kronecker product and $\bm{A}^{(\mathrm{chain})}$ is the $T \times T$ adjacency matrix of an undirected \emph{chain}, or ``bucket brigade,'' network whose $T$ nodes are each adjacent to their nearest neighbors along an undirected chain. In this bucket brigade, $A_{ij}^{(\mathrm{chain})}=1$ for $j=i\pm 1$ and $A_{ij}^{(\mathrm{chain})}=0$ otherwise. 
Although one can choose inter-layer coupling matrices other than $\bm{A}^{(\mathrm{chain})}$ for the inter-layer couplings \drt{\cite{kivela2014,de2016quantifying}} (and much of our approach can be generalized to other choices of coupling), we restrict our attention to nearest-neighbor coupling of layers.

It is also tempting to directly apply a standard eigenvector-based centrality \drt{calculation} to the supra-adjacency matrix ${\mathbb{A}}$ by treating it just like any other adjacency matrix despite its \drt{special} structure. However, such an approach neglects to respect the fundamental distinction between intra-layer edges and inter-layer edges that \drt{arises} from the block-diagonal structure of ${\mathbb{A}}$. That is, in such an approach, one treats the inter-layer couplings (i.e., identity arcs) just like any other edge. In general, however, one needs to be careful when studying a temporal network using \drt{a} supra-adjacency matrix formalism because many basic network properties---some of which carry strong implications about a \drt{time-independent} network (e.g., its \drt{spectrum, connectedness properties, and so on})---do not \drt{carry over naturally} without modification to the supra-adjacency matrix.  This issue was discussed for multilayer networks more generally in Refs.~\cite{cozzo2013,de2013b,kivela2014} and more recently in Ref.~\cite{deford2015}.

As a concrete example, \drt{let's examine} hub and authority centralities (i.e., HITS \cite{kleinberg1999}) for a directed temporal network using the supra-adjacency matrix in Eq.~\eqref{eq:supra} by simply inserting it in place of a \drt{time-independent} adjacency matrix in the standard formulas. In other words, we define the hub and authority matrices as $\mathbb{A}{\mathbb{A}}^T$ and $\mathbb{A}^T{\mathbb{A}}$, respectively. At a glance, by noting that the inter-layer couplings are undirected but that the intra-layer edges are directed, we already see that it is not clear whether standard interpretations of hub and authority rankings are still sensible. Nevertheless, one can try this approach for computing generalized hub and authority scores as the dominant eigenvectors of the symmetric matrices ${\mathbb{A}}{\mathbb{A}}^T$ and ${\mathbb{A}}^T{\mathbb{A}}$. The simplicity of this approach makes it pleasing (and tempting), and the two symmetric matrices do have a block structure. However, in contrast to $\mathbb{A}$ (whose blocks on and off of the main diagonal encode intra-layer and inter-layer edges, respectively), the blocks in the matrices ${\mathbb{A}}{\mathbb{A}}^T$ and ${\mathbb{A}}^T{\mathbb{A}}$ no longer separate neatly into describing only a single type of edge (i.e., inter-layer versus intra-layer edges).
 
The problem with this construction becomes particularly clear in the limit of strong inter-layer coupling (i.e., as $\omega\to\infty$), for which $\mathbb{A}\approx\omega(\bm{A}^{(\mathrm{chain})}\otimes\mathbf{I})$. Because $\bm{A}^{(\mathrm{chain})}$ is symmetric, it follows that $\mathbb{A}{\mathbb{A}}^T\approx\mathbb{A}^T{\mathbb{A}}\approx\omega^2( \bm{A}^{(\mathrm{chain})} (\bm{A}^{(\mathrm{chain})})^T \otimes \mathbf{I})$. Unfortunately, it is useless to compute hub and authority scores of an undirected chain. Specifically, the corresponding hub/authority centrality matrix (whose dominant eigenvector gives the hub/authority scores) of the undirected \drt{chain} becomes
\begin{equation}
 	\bm{A}^{(\mathrm{chain})}(\bm{A}^{(\mathrm{chain})})^T = (\bm{A}^{(\mathrm{chain})})^T\bm{A}^{(\mathrm{chain})} = \left[ \begin{array}{ccccccc}
 1 & 0 & 1 & 0 & \cdots & &\\ 
 0 & 2 & 0 & 1 & \ddots & &\\ 
 1 & 0 & 2 & 0 &\ddots & 0 &\\
 0 & 1 & 0 & 2 &\ddots & 1 & 0 \\
 \vdots & \ddots &\ddots&\ddots&\ddots & 0 & 1\\
 &&0&1&0&2&0\\
 &&&0&1&0&1\\
 \end{array}
	 \right]\,,
 \label{eq:chain_bad}
\end{equation}
revealing that the hub/authority scores of the \drt{even-indexed} and odd-indexed nodes \drt{decouple} from each other. 
The resulting matrix is no longer irreducible, which can \drt{lead to nonuniqueness} of dominant eigenvectors and/or can also cause the entries of a dominant eigenvector to be identically $0$ for a large number of nodes.\footnote{By inspection, the matrix in Eq.~\eqref{eq:chain_bad} is not irreducible, so we cannot apply the Perron--Frobenius theorem for \drt{irreducible} nonnegative matrices \cite{meyer}. \drt{There exist many variations of Perron--Frobenius theory, including ones that are applicable to reducible matrices (e.g., see Sec.~8.3.1 of \cite{meyer}), which one can use to study phenomena that arise in the absence of irreducibility. In our case, we find two types of scenarios,} depending on whether $N$ is odd or even. For even $N$, the largest eigenvalue of $\bm{A}^{(\mathrm{chain})}(\bm{A}^{(\mathrm{chain})})^T $ has a multiplicity of two and a corresponding two-dimensional eigenspace that is spanned by vectors in which either the \drt{even-indexed} or odd-indexed entries are $0$. \drt{Consequently, there is not a unique dominant eigenvector, so there is not a unique ranking of nodes.} For odd $N$, there is \drt{a single} dominant eigenvalue; however, its eigenvector has entries that are identically $0$ for even-indexed nodes, so only half of the nodes are ranked in a nontrivial way. \label{foot:unstable_evecs}} 
 Both issues are detrimental if one wants to rank nodes based on some notion of importance. For example, for large values of $\omega$, we observe oscillations and numerical instabilities when attempting to generalize hub and authority centralities in this way.

\clearpage
\section{Temporal Coupling of Eigenvector-Based Centralities}\label{sec:method}

In this section, we present a mathematical formalism for eigenvector-based centralities in temporal networks that \drt{treats} inter-layer and intra-layer edges as \emph{distinct} types of edges and ensures appropriate behavior for all $\omega > 0$. Similar to prior investigations using multilayer representations of temporal networks \cite{mucha2010,kivela2014}, we seek to develop an approach that \drt{involves} neither a heuristic averaging of centralities from individual layers nor \drt{invokes} the centrality for a single network obtained from the aggregation of network layers (e.g., summing the network edges across time).

The remainder of this section is organized as follows. In Sec.~\ref{sec:main}, we present our methodology for temporal eigenvector-based centrality in terms of the dominant eigenvector of a {\it supra-centrality matrix}. In Sec.~\ref{sec:joint}, we introduce the concepts of \drt{\emph{joint}, \emph{marginal}, and \emph{conditional centrality}, and we use them} to study decoupled centralities of nodes and layers based on the centralities of node-layer pairs. In Sec.~\ref{sec:toy}, we illustrate these concepts for an example synthetic network.

\subsection{Inter-Layer Coupling of Centrality Matrices}\label{sec:main}

To avoid the problems that arise from ignoring the distinction between inter-layer edges and intra-layer edges, we define a somewhat more nuanced generalization of eigenvector-based centralities. To preserve the special role of inter-layer edges, we directly couple the matrices that define the eigenvector-based centrality measure within each temporal layer (e.g., ordinary adjacency matrices for eigenvector centrality). That is, one can cast any eigenvector-based centrality in terms of some matrix $\mathbf{C}$ that is a function of the adjacency matrix $\mathbf{A}$. For example, hub and authority scores are the leading eigenvectors of the matrices $\mathbf{A}\mathbf{A}^T$ and $\mathbf{A}^T\mathbf{A}$, respectively (using the convention that elements $A_{ij}$ indicate $i\to j$ edges). Letting $\mathbf{C}^{(t)}$ denote the centrality matrix for layer $t$, we couple these centrality matrices with inter-layer couplings of strength $\omega$ in a (rescaled) {\it supra-centrality matrix}
{
\begin{equation}
	\mathbb{{C}}(\epsilon) = \left[ \begin{array}{cccc} 
 \epsilon\mathbf{C}^{(1)} & \mathbf{I} & 0 & \cdots\\ 
 \mathbf{I} & \epsilon\mathbf{C}^{(2)} & \mathbf{I} &\ddots\\ 
 0 & \mathbf{I} & \epsilon\mathbf{C}^{(3)} &\ddots\\
 \vdots & \ddots &\ddots&\ddots\\
 \end{array}
 \right]\,.
 \label{eq:sup_centrality1}
\end{equation}}
~\\
\drt{We are defining the supra-centrality matrix using a scaling factor $\epsilon=1/\omega$, because it gives convenient mathematical notation for our forthcoming singular perturbation analysis. However, because it is more intuitive (and standard) to describe layers as being coupled together by weight $\omega$, we refer to the limit $\epsilon\to 0^+$ (i.e., $\omega\to\infty$) as the strong-coupling regime.}

One can interpret the parameter $\epsilon>0$ as a tuning parameter that controls how strongly a given physical node's centrality is coupled to itself between neighboring time layers. (See the related discussions in \cite{bassett2013,bazzi2015} in the context of multilayer community structure.) That is, the intuition for a specified eigenvector-based centrality proceeds within each individual layer as in the associated centrality's original definition, and the additional inter-layer coupling introduces contributions to centrality from the network structure in neighboring layers. Of particular interest are the limits in which $\epsilon\to\infty$ (i.e., decoupling of layers) and $\epsilon\to0^+$ (i.e., a particular notion of order-preserving aggregation). See the related discussions in \cite{arenas-natphys2013,raddichi-prx2014}. We expect the $\epsilon\to0^+$ limit to yield principled time-averaged centralities of nodes. Note that such a notion reflects the layers having an intrinsic temporal ordering and should in general yield different results from calculating the centralities of \drt{summed} adjacency layers (i.e., directly summing the corresponding entries in these matrices) or from an unweighted averaging of centralities across otherwise uncoupled layers. 

\drt{The notation of Eq.~\eqref{eq:sup_centrality1} assumes implicitly} that every physical node $i$ appears in every time layer $t$ (where $i\in\{1,\dots,N\}$ \drt{and} $t\in\{1,\dots,T\}$). Although this notation is consistent with that used in the development of multilayer modularity \cite{mucha2010}, it is important to call attention to practical issues in treating situations in which some physical nodes do not appear across all layers \cite{kivela2014,de2013b,cozzo2013}. When defining multilayer modularity, there are no difficulties with removing nodes from the layers in which they do not appear as long as the correct identity inter-layer edges are coded appropriately. (Indeed, see the U.S. Senate roll-call voting example of \cite{mucha2010}.) In contrast, for reasons that will become clearer as we develop our singular perturbation analysis in the strong-coupling limit (see Sec.~\ref{sec:TimeAve}), temporal generalization of eigenvector-based centralities in this limit requires that all physical nodes are taken into account across all layers, even when they do not appear in a layer. In other words, one must account for each physical node $i$ as a ``ghost node" in any layer in which $i$ does not appear in the data.  The ghost node is adjacent to its counterparts in neighboring layers via inter-layer edges, and it therefore maintains connectivity to the full multilayer network; however, it does not have any intra-layer edges, as such edges are ``forbidden.''

Before continuing, we briefly comment on assumptions that we make about the supra-centrality matrix $\mathbb{C}(\epsilon)$. In the construction of eigenvector-based centralities for \drt{time-independent} networks, it is typically assumed that a given centrality matrix is nonnegative and irreducible \cite{benzi2013,estrada2010,kleinberg1999,pagerank,langville2006,gleich2014}. Similarly, we assume that $\mathbb{C}(\epsilon)$ is nonnegative and irreducible for any $\epsilon>0$. Our motivation is that the Perron--Frobenius theorem for nonnegative matrices \cite{meyer} ensures that the largest (positive) eigenvalue has multiplicity one and that its corresponding eigenvector is nonnegative and unique, which are both beneficial properties for ranking nodes based on a notion of importance. Similar to the case of \drt{time-independent} networks, one can guarantee that the matrix {$\mathbb{C}(\epsilon)$} is both nonnegative and irreducible by placing simple constraints on the properties of the temporal network. 
For example, consider the matrix {$\mathbb{C}(\epsilon)$} and its associated network---that is, the network in which every nonzero entry of $\mathbb{C}(\epsilon)$ \drt{corresponds to an edge from some node $i$ to some other node $j$, and the weight is given by the value of the entry.} (In practice, one can use such an approach to study any matrix if one interprets its nonzero entries in terms of a network.) It follows that $\mathbb{C}(\epsilon)$ is irreducible and nonnegative if this associated network is strongly connected. A sufficient (but not necessary) condition to assure this is that all centrality matrices $\mathbf{C}^{(t)}$ are nonnegative and the aggregated matrix $\sum_t \mathbf{C}^{(t)}$ itself has an associated network that is strongly connected. For example, when computing eigenvector centrality for an undirected temporal network (i.e., $\mathbf{C}^{(t)}=\mathbf{A}^{(t)}$), this constraint implies that $A_{ij}^{(t)}\ge0$ and that the aggregation $\sum_t \mathbf{A}^{(t)}$ of the adjacency matrices yields an adjacency \drt{matrix with an associated strongly-connected network.}\footnote{There exist both stronger and weaker versions of such a relation between network structure and the dominant eigenspace of the matrices that are associated with a network. If a network is strongly connected, then the largest (positive) eigenvalue of the matrix has a multiplicity of one, and its corresponding eigenvector is guaranteed to be unique and strictly positive. If a network is weakly connected and if all nodes are contained in the union of the largest in-, out-, and strongly-connected components, then the matrix has a largest (positive) eigenvalue with a multiplicity of one, and its corresponding eigenvector is both unique and nonnegative. In other cases, the eigenvector corresponding to the largest eigenvalue may or may not be unique. If \drt{there are edges with negative values,} the eigenvector corresponding to the largest eigenvalue may not be nonnegative.\label{foot:irreducible}} 
\drt{In general, however, the irreducibility of $\mathbb{C}(\epsilon)$ depends on that of the centrality matrices (i.e., $\sum_t \mathbf{C}^{(t)}$), rather than on the adjacency matrices.}

\drt{Our assumption that $\mathbb{C}(\epsilon)$ be irreducible places restrictions both on the set $\{\mathbf{C}^{(t)}\}$ and on our choice for how to couple the layers. 
We choose to couple each time layer $t$ to both layers $t+1$ and $t-1$ (when present), so our temporal extension of eigenvector centrality is non-causal; that is, the centralities are coupled both forward and backward in time. In principle, one could choose other strategies for coupling the layers. Note, however, that a causal strategy in the absence of other features (e.g., one could examine different types of ``teleportation'' \cite{gleich2014}) forces $\mathbb{C}(\epsilon)$ to be reducible, because the centralities of past network layers cannot depend on future layers.}

\subsection{Joint, Marginal and Conditional Centrality for Multilayer Networks}\label{sec:joint}

We study the dominant eigenvector $\mathbbm{v}(\epsilon)$ \drt{of $\mathbb{{C}}(\epsilon)$, with corresponding (and largest positive)} eigenvalue $\lambda_\textrm{max}(\epsilon)$ [i.e, {$\mathbb{{C}}(\epsilon)\mathbbm{v}(\epsilon) = \lambda_\textrm{max}(\epsilon)\mathbbm{v}(\epsilon)$]. }The entries of the dominant eigenvector give the centralities of each node-layer pair $(i,t)$; this represents the centrality of physical node $i$ at time $t$. {The dominant eigenvector $\mathbbm{v}(\epsilon)$ of a supra-centrality matrix in Eq.~\eqref{eq:sup_centrality1} gives the centrality of node-layer pairs. That is, the eigenvalue entry $v_{N(t-1) + i}(\epsilon)$ indicates the centrality of node $i$ at time $t$. Such a joint centrality, whether given by $\mathbbm{v}(\epsilon)$ or any other centrality for node-layer pairs, reflects information about the importances of both the nodes and the layers. We develop a simple formalism to decouple these centralities. For concreteness, we use $\mathbbm{v}(\epsilon)$, but this approach can be applied to any centrality measure of node-layer pairs in a multilayer (e.g., temporal) network \drt{including those  not based on eigenvectors. }

Our approach is inspired by multivariate statistics: we define ``joint'', ``marginal'', and ``conditional'' centralities. Joint centrality describes the importances of node-layer pairs, marginal centrality describes the uncoupled centrality of either nodes or layers, and conditional centrality describes the importance of a node-layer pair as compared to, for example, other node-layer pairs \drt{in} that same layer.

To proceed, it is convenient to map the vector $\mathbbm{v}(\epsilon)$, which \drt{has} length $NT$, to an $N\times T$ matrix $\mathbf{W}$, which we define entry-wise by 
\begin{equation}
	W_{it}  = v_{N(t-1) + i}(\epsilon)\,. \label{eq:joint}
\end{equation}
The scalar $W_{it}$ gives the joint centrality of the node-layer pair ($i,t)$; that is, it indicates the centrality of node $i$ at time $t$. We define the \emph{marginal node centrality} (MNC) $x_i$ and \emph{marginal layer centrality} (MLC) $y_t$ by 
\begin{equation}
	x_{i} = \sum_{t}W_{it}\,, \qquad y_{t} = \sum_{i}W_{it}\,.
\label{eq:marginal}
\end{equation}
The values $\{x_i\}$ and $\{y_t\}$ indicate the importances of nodes and layers, respectively, for a particular choice of $\epsilon$. Although we use the summation to compute marginal node and layer centralities, one can also consider other aggregation methods.
We define the \emph{conditional centrality} of node-layer pair $(i,t)$, conditioned on layer $t$, by
\begin{equation}
	Z_{it} = W_{it}/y_t\,.
\label{eq:conditional}
\end{equation}
The scalar $Z_{it}$ indicates the importance of physical node $i$ relative to other physical nodes in layer $t$. For some applications, it can be beneficial to similarly study the conditional centrality of layers conditional on a given node, but we do not explore this notion in the present paper. For a given node $i \in \{1, \dots, N\}$ and time $t\in\{1,\dots,T\}$, \drt{the sets $\{W_{it}\}$ and $\{Z_{it}\}$ of centrality values indicate \emph{trajectories}} for how the \drt{importance} of physical node $i$ changes through time. {We interpret conditional centrality trajectories as follows:} for a given physical node $i$, we study a sequence of centralities in which the $t$th term indicates a centrality that is relative to centralities of node-layer pairs at time $t$. This contrasts with the joint node-layer \drt{centralities;} because $\|\mathbbm{v}(\epsilon)\|_2=1$, joint node-layer centralities \drt{indicate a notion of importance} that is relative to all node-layers pairs. 
}

\begin{figure}[t!]
\centering
\includegraphics[width=.9\linewidth]{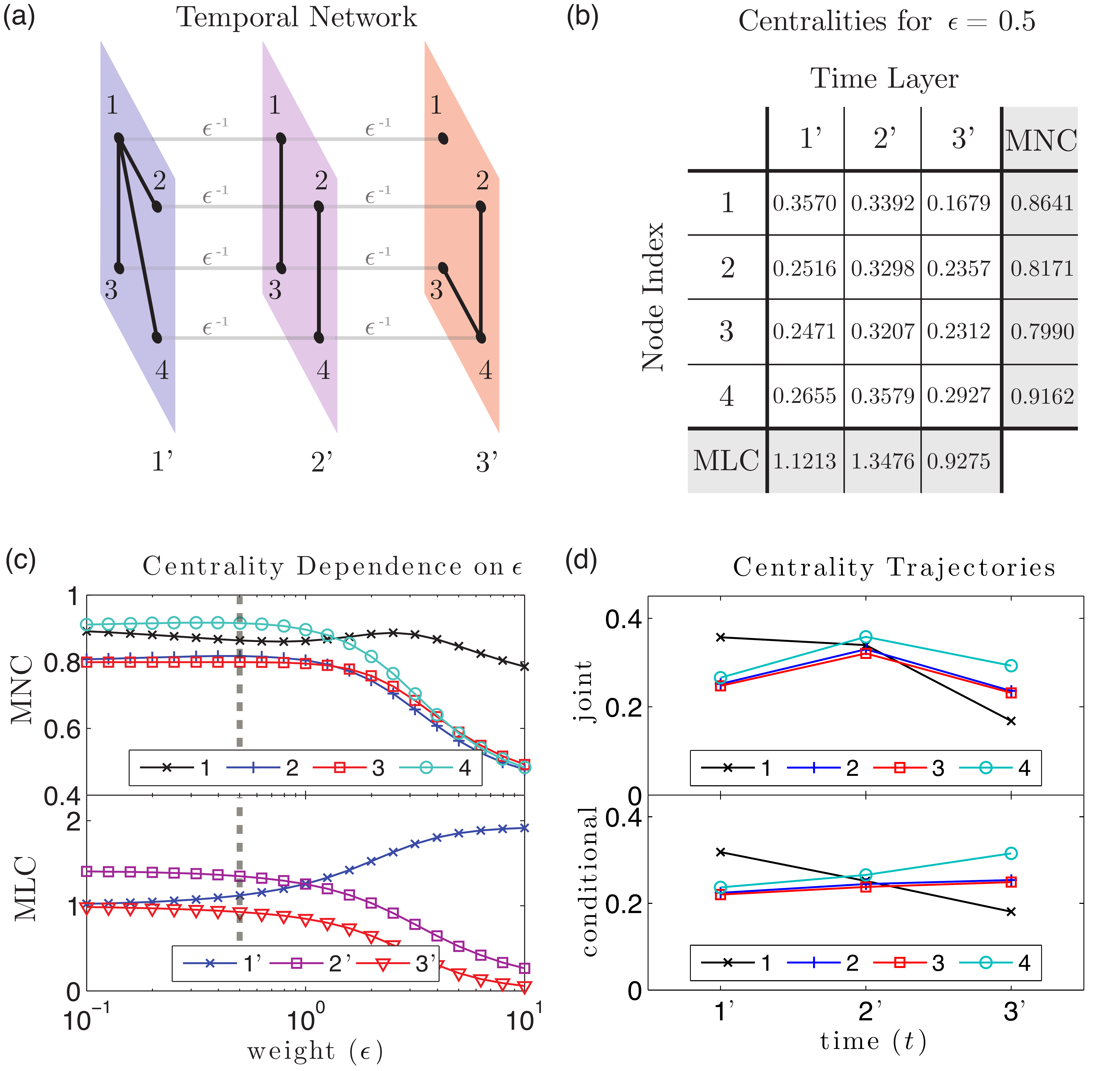}
\caption{
Eigenvector centralities for an example undirected temporal network, with $N=4$ nodes and $T=3$ layers, in which we use the layers' adjacency matrices as the centrality matrices. 
(a)~A temporal network with intra-layer edges (black lines) and inter-layer identity edges (gray lines). 
(b) For $\epsilon=0.5$, we \drt{indicate the joint node-layer centralities \drt{$\{W_{it}\}$} (white boxes), which correspond to the entries in $\mathbbm{v}(\epsilon)$ and represent} the centralities of each physical node in each layer (there are $N\times T$ such centralities). We refer to the marginal centralities \drt{$\{x_{i}\}$ and $\{y_t\}$} (shaded boxes) that one obtains by summing rows and columns, respectively, as the ``marginal node centralities'' (MNC) and ``marginal layer centralities'' (MLC).
(c) 
MNC and MLC versus the coupling parameter $\epsilon$. For small $\epsilon$, the dominant time layer is the central layer (2'); for large $\epsilon$, the dominant layer is layer 1', which contains the centrality matrix with the dominant
(i.e., largest positive) eigenvalue.
The vertical dashed line indicates the value $\epsilon=0.5$ that we use in panels (b) and (d). 
(d) We examine the nodes' centrality trajectories by plotting the joint node-layer centralities (upper subpanel) and conditional node-layer centralities (lower subpanel) versus time.
}
\label{fig:toy1}
\end{figure}

\subsection{Synthetic Temporal Network Example}\label{sec:toy}

We illustrate the concepts of joint, marginal, and conditional centrality with a toy example. The network, which we show in Fig.~\ref{fig:toy1}(a), consists of $N=4$ physical nodes and $T=3$ time layers. We compute eigenvector centralities of this (undirected) temporal network by setting each layer's centrality matrix to be its adjacency matrix and solving the dominant-eigenvalue equation, $\mathbb{C}(\epsilon)\mathbbm{v}(\epsilon)=\lambda_\textrm{max}(\epsilon)\mathbbm{v}(\epsilon)$, for several choices of $\epsilon$. In Fig.~\ref{fig:toy1}(b), we summarize the centrality measures for $\epsilon=0.5$ in a matrix. The entry in row $i$ and column $t$ gives {$W_{it}$. The MNCs $\{x_i\}$ and the MLCs $\{y_t\}$ are in the shaded boxes, and they indicate the relative centralities of the physical nodes and layers, respectively, for the chosen value of $\epsilon$. 

In Fig.~\ref{fig:toy1}(c), we plot the dependence of the MNC (upper subpanel) and MLC (lower subpanel) on \drt{the} coupling strength $\epsilon$. For small $\epsilon$, the dominant time layer is layer 2'; for large $\epsilon$, the dominant layer is layer 1'. This is unsurprising: when considering the centrality matrices of the layers in isolation, the dominant eigenvalue of the centrality matrix of layer 1' is larger than that for the other layers. The choice of $\epsilon$ is very important, and centralities can depend discontinuously on $\epsilon$ because of eigenvalue crossings. In this example, we obtain three qualitatively different regimes: (i)~a strong-coupling regime in which the centralities are similar to what we expect in the limit $\epsilon\to0^+$; (ii) a weak-coupling regime in which the centralities behave similarly to what they do in the $\epsilon\to\infty$ limit; and (iii) an intermediate-coupling regime in which the centralities transition between these two limiting cases. (Compare this result to the phase-transition phenomena for graph Laplacians of multilayer networks discussed in Refs.~\cite{arenas-natphys2013,raddichi-prx2014}.) We also explore these regimes for our case study with the MGP network in Sec.~\ref{sec:mgp}.

In Fig.~\ref{fig:toy1}(d), we plot joint node-layer centralities (upper subpanel) and the conditional node-layer centralities (lower subpanel) that correspond to the four physical nodes. We show results for $\epsilon=0.5$. Note that the conditional node-layer centralities for physical node 4 increase over time, whereas they decrease for physical node 1. This is sensible for the temporal network in Fig.~\ref{fig:toy1}(a): at time $t=1$', physical node 1 has the largest degree and physical node 4 has the smallest degree; however, at time $t=3$', physical node 1 has the smallest degree and physical node 4 has the largest degree.

\section{Singular Perturbation in the Strong-Coupling Limit}\label{sec:TimeAve}

Given the joint node-layer centralities and conditional node-layer centralities that correspond to a physical node $i$, it is possible to define a notion of ``time-averaged centrality'' by summing one of these centralities across the layers (e.g., the prior yields the MNC). However, it is not clear which is preferable, and these centralities are sensitive to the value of $\epsilon$.  Alternatively, \drt{one} can define a time-averaged centrality by studying the limit $\epsilon\to0^+$. In this limit, the conditional node-layer centrality of every physical node $i$ \drt{is} constant across the time layers. 

Examining centralities as $\epsilon\to0^+$ provides a principled approach for calculating time-averaged centralities. However, the supra-centrality matrix $\mathbb{C}(\epsilon)$ given by Eq.~\eqref{eq:sup_centrality1} becomes singular at $\epsilon=0$, which complicates the consideration of this limit. The intra-layer connectivity is completely eliminated, and the network decomposes into $N$ connected components. That is, the matrix is no longer irreducible, so the Perron--Frobenius theorem no longer holds. Indeed, at $\epsilon=0$, the dominant eigenvalue has an $N$-dimensional eigenspace. In contrast, for $\epsilon>0$, the dominant eigenvalue has a single eigenvector.

To overcome this issue, we derive a singular perturbation expansion in the limit $\epsilon\to0^+$. In Sec.~\ref{sec:zero}, we further explore the singularity that arises in the strong-coupling limit. In Secs.~\ref{sec:perturb} and \ref{sec:movers}, we give \drt{zeroth-order} and first-order perturbation expansions, which lead to principled expressions for time-averaged centralities and first-order-mover scores, respectively. We give higher-order expansions in an appendix.  
 {In Sec.~\ref{sec:algorithm}, we summarize our procedure and discuss the computational complexity of computing time-averaged centralities and first-order-mover scores. }

\subsection{Singularity at Infinite Inter-Layer Coupling}\label{sec:zero}

In this section, we develop a perturbation analysis of the dominant eigenspace (i.e., the eigenspace of the largest \drt{positive} eigenvalue) of $\mathbb{C}(\epsilon)$ [see Eq.~\eqref{eq:sup_centrality1}] in the limit $\epsilon\to0^+$. \drt{To allow this analysis to have more broad applicability,} we do a perturbation expansion using the following \drt{general} form of coupling of block matrices:
\begin{equation}
	\mathbb{M}(\epsilon)  = \mathbb{B} +  \epsilon \mathbb{G} \,,
\label{eq:sup_centrality2}
\end{equation}
\drt{where $\mathbb{G} = \text{diag}[ \mathbf{M}^{(1)},\dots, \mathbf{M}^{(T)}]$, $\mathbb{B}=\bm{A}\otimes \bf{I}$, and the $T\times T$ matrix $\bm{A}$ {(recall Table \ref{table:notation}) encodes the inter-layer coupling, where entry $A_{tt'}$ indicates} how layer $t$ is coupled to layer $t'$.} We recover the supra-centrality matrix $\mathbb{C}(\epsilon)$ in Eq.~\eqref{eq:sup_centrality1} by using nearest-neighbor layer coupling, $\bm{A} = \bm{A}^{(\textrm{chain})}$, and centrality matrices along the diagonal (i.e., $\mathbf{M}^{(t)}=\mathbf{C}^{(t)}$). Additionally, similar to our assumptions for Eq.~\eqref{eq:sup_centrality1}, we assume that $\mathbb{M}(\epsilon) $ is nonnegative and irreducible for any $\epsilon > 0 $. These assumptions hold as long as the summation of matrices ($\sum_t\mathbf{M}^{(t)}$) and inter-layer coupling matrix $\bm{A}$ each correspond to a strongly connected network (see footnote 3 in Sec.~\ref{sec:joint}).

\drt{We begin by studying the dominant eigenspace for the matrix $\mathbb{M}(\epsilon)$ at $\epsilon=0$. We thus consider the matrix
\begin{align}
	\mathbb{M}(0)  &=  \bm{A} \otimes \mathbf{I}\,,
\end{align} 
and we will show that the dominant eigenspace of $\mathbb{M}(0)$ is $N$-dimensional, implying that one cannot obtain the unique dominant \drt{eigenvector} of $\mathbb{M}(\epsilon)$ for the $\epsilon\to 0^+$ limit simply by setting $\epsilon=0$. Instead, one needs to do a singular perturbation analysis.}
To facilitate our discussion, \drt{we define} an $NT\times NT$ {stride permutation matrix $\mathbb{P}$ \cite{golub}} with entries $P_{kl} = 1$ for $l=\lceil k/N\rceil+T\,[(k-1)\bmod N]$ and $P_{kl} = 0$ \drt{otherwise. (Recall that $\lceil\cdot\rceil$ denotes the ceiling function.)} Note that $\mathbb{P}$ permutes node-layer \drt{indices, so we can easily go back and forth} between ordering the node-layer pairs \drt{first} by time and then by physical node index, or vice versa (i.e., ordering them \drt{first} by physical node index and then by time).
\drt{The main benefit of defining $\mathbb{P}$ is that one can express $\mathbb{M}(0)$ in the form $\mathbb{M}(0) =  \mathbb{P}  \left( \mathbf{I} \otimes \bm{A}  \right)\mathbb{P}^T$ \cite{demmel1997applied}, where}
\begin{equation}
	\mathbf{I} \otimes \bm{A}  = 
\left[ \begin{array}{cccc} 
 \bm{A} & 0 & 0 & \cdots\\ 
0& \bm{A}  & 0& \\ 
0& 0 &\bm{A} &\ddots\\
 \vdots &   &\ddots&\ddots\\
 \end{array}
 \right] \, .
\label{eq:perm}
\end{equation}
\drt{Clearly, $\mathbf{I} \otimes \bm{A}$} decouples into $N$ identical eigenvalue equations for the inter-layer coupling matrix $\bm{A}$. 
\drt{Because $\mathbb{P}$ is a unitary matrix, this decoupling also occurs for $\mathbb{M}(0)$.
Specifically,} the $NT$ eigenvalues of $\mathbf{I} \otimes \bm{A}$ are given by the $T$ eigenvalues of $\bm{A}$, where \drt{each eigenvalue has multiplicity $N$} and a corresponding $N$-dimensional eigenspace spanned by the vectors \drt{that can be constructed using} the eigenvectors of $\bm{A}$ \drt{(i.e.,} with appended $0$ values in appropriate coordinates). 

\drt{We explain this construction in more detail for the dominant eigenspace.} \drt{Let} {$\nu$} denote the largest \drt{positive} eigenvalue of $\bm{A}$, and let $\bm{u} =[u_1,\dots,u_T]^T $ denote its corresponding eigenvector. \drt{Because of the block-diagonal structure of Eq.~\eqref{eq:perm}, the largest eigenvalue $\lambda_0$ of $\mathbf{I} \otimes \bm{A}$ is $\lambda_0=\nu $, and its corresponding eigenspace is spanned by the $N$ eigenvectors $\{\mathbbm{u}_i\}$, where $\mathbbm{u}_i = [\bm{0}^T,\dots,\bm{0}^T,\bm{u}^T,\bm{0}^T,\dots,\bm{0}^T]^T$.}
That is, the $i$th block of $\mathbbm{u}_i$ is given by $\bm{u}$, and all of the other blocks are vectors of \drt{zeros}.
Consequently, one can obtain the $N$ dominant eigenvectors of $\mathbb{M}(0) =  \bm{A}  \otimes \mathbf{I}$ using the permutations $\{\mathbb{P}\mathbbm{u}_i\}$. That is, they have the general form $\sum_j \alpha_j\mathbb{P}\mathbbm{u}_j$, where the constants $\{\alpha_i\}$ must satisfy $\sum_i\alpha_i^2=1$ to ensure that the vector is normalized. 
\drt{Because there does not exist a unique dominant eigenvector of $\mathbb{M}(0)$ \drt{since its dominant eigenspace is $N$ dimensional}, we need to develop a singular perturbation analysis to obtain a unique solution for the dominant eigenvector of $\mathbb{M}(\epsilon)$ in the $\epsilon\to0^+$ limit.}

\drt{When} the network layers are coupled by an undirected chain network, \drt{the} inter-layer coupling matrix is given by $\bm{A}=\bm{A}^{(\textrm{chain})}$, which has $N$ eigenvalues and eigenvectors given by \cite{lovasz1973}
\begin{align}
	\nu^{(\textrm{chain})} &= 2\cos\left({\frac{n\pi}{T+1}}\right)\,,\label{eq:chain_1}\\
\bm{u}^{(\textrm{chain})} &= \frac{1}{\sqrt{\gamma_n}} \left[\sin\left( \frac{n\pi}{T+1}\right),\sin\left( \frac{2n\pi}{T+1}\right),\dots,\sin\left(  \frac{Tn\pi }{T+1} \right)\right]^T\,,
	\label{eq:chain_2}
\end{align}
where \drt{$n\in\{1,...,N\}$ and} the normalization constant is $\gamma_n = \sum_{t=1}^T\sin^2\left[ {n\pi t}/{(T+1)}\right]$. Setting $n=1$ gives the dominant eigenvalue and its corresponding eigenvector..

\subsection{Zeroth-Order Expansion and Time-Averaged Centrality}\label{sec:perturb}

In this section, we study the zeroth-order expansion of the dominant eigenvector $\mathbbm{v}(\epsilon)$ in the limit $\epsilon\to0^+$.  
As we \drt{now} show, the conditional node-layer centralities \drt{$\{Z_{it}\}$} corresponding to a given physical node $i$ become constant across time in this limit. \drt{(Recall our definitions of joint, marginal and conditional centralities in Sec.~\ref{sec:joint}.)} \drt{For each physical node,} we refer to \drt{this limiting} value as \drt{its} \emph{time-averaged centrality}. By examining the first-order expansion \drt{of $\mathbbm{v}(\epsilon)$ in the limit $\epsilon\to0^+$}, we show in Eq.~\eqref{eq:ap1} that one can obtain \drt{the time-averaged centralities} as \drt{the eigenvector components corresponding to the largest eigenvalue} of a matrix of size $N\times N$.

We consider the \drt{dominant-eigenvalue} equation
\begin{equation}
	\lambda_\textrm{max}(\epsilon) \mathbbm{v}(\epsilon) = \mathbb{M}(\epsilon)\mathbbm{v}(\epsilon) = \mathbb{B}\mathbbm{v}(\epsilon)+\epsilon \mathbb{G}\mathbbm{v}(\epsilon) \,. \label{eq:eval1}
\end{equation}
We expand $\lambda_\textrm{max}(\epsilon)$ and $\mathbbm{v}(\epsilon)$ for small $\epsilon$ by writing $\lambda_\textrm{max}(\epsilon) = \lambda_0+\epsilon\lambda_1 + \cdots$ and $\mathbbm{v}(\epsilon) = \mathbbm{v}_0 + \epsilon \mathbbm{v}_1 +  \cdots $ to obtain $k$th-order approximations:
$\lambda_\textrm{max}(\epsilon) \approx \sum_{j=0}^k\epsilon^j\lambda_j$ and $\mathbbm{v}(\epsilon) \approx \sum_{j=0}^k\epsilon^j\mathbbm{v}_j$. 
\drt{We use superscripts to indicate powers of $\epsilon$ in the terms in the expansion, and we use subscripts for the terms that are multiplied by the power of $\epsilon$.}
\drt{Note that $\lambda_0$ and $\mathbbm{v}_0$ respectively} indicate the dominant eigenvalue and corresponding eigenvector in the limit $\epsilon\to0^+$. 
\drt{Successive terms in these expansions represent higher-order derivatives of $\lambda_\textrm{max}(\epsilon)$ and $\mathbbm{v}(\epsilon)$, and each term assumes appropriate smoothness of these functions. }

Our strategy is to develop consistent solutions to Eq.~\eqref{eq:eval1} for increasing values of $k$. Starting with the first-order \drt{approximation}, we substitute $\lambda_\textrm{max}(\epsilon) \approx \lambda_{0} + \epsilon \lambda_{1}$ and $\mathbbm{v}(\epsilon) \approx \mathbbm{v}_0 + \epsilon \mathbbm{v}_1 $ into Eq.~\eqref{eq:eval1} and collect the \drt{zeroth-order} and first-order terms in $\epsilon$ to obtain
\begin{align}
	\left(\lambda_0\mathbb{I} - \mathbb{B}\right)\mathbbm{v}_0 &= 0 \,, \label{eq:first_1} \\
\left(\lambda_0\mathbb{I} - \mathbb{B}\right)\mathbbm{v}_1 &= \left(\mathbb{G} - \lambda_1\mathbb{I}\right)	\mathbbm{v}_0 \label{eq:first_2} \,,
\end{align}
where $\mathbb{I}$ is the $NT\times NT$ identity matrix. Equation~\eqref{eq:first_1} is exactly the system that we studied in Sec.~\ref{sec:zero} [see Eq.~\eqref{eq:sup_centrality2} with $\epsilon=0$], where we found that the operator $\lambda_0\mathbb{I} - \mathbb{B}$ is singular and has an $N$-dimensional null \drt{space. (This is the dominant eigenspace of $\mathbb{B}$.)} We also found that Eq.~\eqref{eq:first_1} has a general solution of the form
\begin{equation}
	{\lambda_0 = \nu,} \qquad \mathbbm{v}_0=\sum_j\alpha_j \mathbb{P}\mathbbm{u}_j \,,  \label{eq:v0}
\end{equation}
where $\{\alpha_i\}$ are constants that satisfy the constraint that $\mathbbm{v}_0$ has magnitude $1$ (i.e., $\sum_i \alpha_i^2=1$). We defined $\mathbbm{u}_i$ just below Eq.~\eqref{eq:perm}.

To find the set $\{\alpha_i\}$ of unique constants that determine $\mathbbm{v}_0$, we \drt{seek} a solvability condition in the first-order terms. Using the fact that the null space of $\lambda_0\mathbb{I} - \mathbb{B}$ is $\textrm{span}(\mathbb{P}\mathbbm{u}_1,\dots, \mathbb{P}\mathbbm{u}_N)$ for any physical node $i$, it follows that $(\mathbb{P}\mathbbm{u}_i)^T\left(\lambda_0\mathbb{I} - \mathbb{B}\right)\mathbbm{v}_1 = 0$, and left-multiplying Eq.~\eqref{eq:first_2} by $(\mathbb{P}\mathbbm{u}_i)^T$ leads to
\begin{align}
	\mathbbm{u}_i^T\mathbb{P}^T \mathbb{G} \mathbbm{v}_0  &=  \lambda_1 \mathbbm{u}_i^T\mathbb{P}^T \mathbbm{v}_0 \,.\label{eq:base0}
\end{align}
Using the solution of $\mathbbm{v}_0$ in Eq.~\eqref{eq:v0}, we obtain 
\begin{align}
	\sum_j\alpha_j \mathbbm{u}_i^T\mathbb{P}^T \mathbb{G}  \mathbb{P}\mathbbm{u}_j  &=  \lambda_1 \sum_j\alpha_j  \mathbbm{u}_i^T\mathbb{P}^T \mathbb{P}\mathbbm{u}_j = \lambda_1 \alpha_i\,,\label{eq:eig_a0}
\end{align}
because $\mathbb{P}^T\mathbb{P}=\mathbb{P}\mathbb{P}^T=\mathbb{I}$ and $ \mathbbm{u}_i^T \mathbbm{u}_j=\delta_{ij}$, where $\delta_{ij}$ is the Kronecker delta. Letting $\bm{\alpha}= [\alpha_1,\dots,\alpha_N]^T$, Eq.~\eqref{eq:eig_a0} corresponds to an $N$-dimensional eigenvalue equation, 
\begin{equation}
	\mathbf{X}^{(1)} \bm{\alpha} =  \lambda_1 \bm{\alpha}\,,\label{eq:ap1}
\end{equation}
where the matrix $\mathbf{X}^{(1)}$ has elements
\begin{equation}
	{{X}^{(1)}_{ij} = \mathbbm{u}_i^T\mathbb{P}^T \mathbb{G}  \mathbb{P}\mathbbm{u}_j =  \sum_{t} {M}^{(t)}_{ij} u_t^2\,.\label{eq:M1star_0}}
\end{equation} 
Our assumption that $\mathbb{M}(\epsilon)$ is nonnegative and irreducible for any $\epsilon>0$ ensures that $\mathbf{X}^{(1)}$ is also nonnegative and irreducible. By the Perron--Frobenius theorem for nonnegative matrices \cite{meyer}, the largest \drt{positive} eigenvalue $\lambda_1$ of $\mathbf{X}^{(1)}$ has a multiplicity of one, and its eigenvector $\bm{\alpha}$ is unique and has nonnegative entries. (See Sec.~\ref{sec:main} \drt{and footnote \ref{foot:irreducible}.}) We normalize the solution $\bm{\alpha}$ to Eq.~\eqref{eq:ap1} by $\sum_i {\alpha}^2_i=1$ and substitute the normalized solution into Eq.~\eqref{eq:v0} to obtain the zeroth-order term $\mathbbm{v}_0$.

{When the layers are coupled by an undirected chain and the block matrices are the layers' centrality matrices (i.e., $\mathbf{M}^{(t)}=\mathbf{C}^{(t)}$), we obtain}
\begin{equation}
	{X}^{(1)}_{ij} = \gamma_1^{-1}\sum_{t} {C}^{(t)}_{ij} \sin^2 \left(\frac{\pi t}{T+1}\right)\,, \label{eq:M1star}
\end{equation} 
where $\gamma_1 = \sum_{t=1}^T\sin^2\left( {\pi t}/{(T+1)}\right)$ is the normalization constant for the dominant eigenvector $\bm{u}^{(\textrm{chain})}$ given by $n=1$ in Eq.~\eqref{eq:chain_2}. 
{In this case, recall} that the vector $\mathbbm{v}_0$ is the dominant eigenvector of $\mathbb{C}(\epsilon)$ in the limit $\epsilon\to0^+$ and gives the joint node-layer centralities \drt{in} this limit. By inspection, \drt{we see that} the elements of $\mathbbm{v}_0$ are $\alpha_i\sin(\pi t/(T+1))$ for node-layer pair $(i,t)$. \drt{Because these correspond to the limiting $\epsilon\to0^+$ entries of vector $\mathbbm{v}(\epsilon)$, they are independent of $\epsilon$. At the same time,} conditional centrality of node-layer pair $(i,t)$ is $\alpha_i$ (up to a normalization constant), independent of the layer $t$. {That is, the conditional node centrality trajectories become constant across time in the limit $\epsilon\to0^+$.}
\drt{These values of $\{\alpha_i\}$} arise naturally from our perturbative expansion in the supra-centrality framework, independently of the value of $\epsilon$. By contrast, recall that the marginal node centralities \drt{(MNCs)} reflect averaging the joint centralities across time layers for a specific choice of $\epsilon$. Accordingly, we hereafter refer to the entry ${\alpha}_i$ in the vector $\bm{\alpha}$ as the {\emph{time-averaged node centrality} of physical node $i$. 
{Because our approach can also be applied to multilayer networks that are not necessarily temporal, we call ${\alpha}_i$ the \emph{layer-averaged node centrality} for these situations.}

\subsection{First-Order Expansion and First-Order-Mover Scores}\label{sec:movers}

In this section, we show that the first-order expansion of Eq.~\eqref{eq:eval1} leads to a linear system [see Eq.~\eqref{eq:betas3}], which we solve to obtain a measurement of the variation over time of each physical node's centrality trajectory [see Eq.~\eqref{eq:fmover}]. Specifically, as one increases $\epsilon$ above $0^+$, the first-order expansion, {(which \drt{accounts for first} derivatives with respect to $\epsilon$)}, captures the dominant changes in centrality trajectories for small values of $\epsilon$.  (In \drt{the} appendix, we derive expressions for higher-order \drt{terms that account for higher-order derivatives with respect to $\epsilon$}.)

In Sec.~\ref{sec:perturb}, we derived closed-form expressions for $\lambda_0$ and $\mathbbm{v}_0$ and an eigenvalue equation satisfied by $\lambda_1$. We now solve for $\mathbbm{v}_1$ to complete our first-order approximation. For notational convenience, we define $\mathbb{L}_0 = \lambda_0\mathbb{I}-\mathbb{B}$ and $\mathbb{L}_1 = \mathbb{G} - \lambda_1\mathbb{I}$, so Eq.~\eqref{eq:first_2} becomes $\mathbb{L}_0\mathbbm{v}_1=\mathbb{L}_1\mathbbm{v}_0$. Letting $\mathbb{L}_0^\dagger$ denote the Moore--Penrose pseudoinverse of $\mathbb{L}_0$, 
we write
\begin{equation}
	\mathbbm{v}_1 
= \mathbb{L}_0^{\dagger}\mathbb{L}_1\mathbbm{v}_0 + \sum_j \beta_j \mathbb{P}\mathbbm{u}_j 
	= \mathbb{L}_0^{\dagger}\mathbb{G}\mathbbm{v}_0 + \sum_j \beta_j \mathbb{P}\mathbbm{u}_j\,. \label{eq:v1}
\end{equation}
We simplify the first term in Eq.~\eqref{eq:v1} using $\mathbb{L}_1=\mathbb{G}-\lambda_1\mathbb{I}$ and $\mathbbm{v}_0 = \sum_j \alpha_j \mathbb{P}\mathbbm{u}_j$ and \drt{noting} that each vector $\mathbb{P}\mathbbm{u}_j$ lies in the null space of each of the matrices $\mathbb{L}_0$ and $\mathbb{L}_0^\dagger$. The second term in Eq.~\eqref{eq:v1} accounts for the projection of $\mathbbm{v}_1$ onto the null space of $\mathbb{L}_0$, where the constants $\beta_i=(\mathbb{P}\mathbbm{u}_i)^T\mathbbm{v}_1$ indicate the projections onto the spanning vectors of the null space. To ensure numerical stability and computational efficiency in practice, we calculate $\mathbb{L}_0^\dagger$ using the identity\drt{\cite{demmel1997applied}}
\begin{equation}
	\mathbb{L}_0^\dagger  = \left(\lambda_0\mathbb{I} - \bm{A} \otimes \mathbf{I} \right)^\dagger  
	= \left(\lambda_0\bm{I} - \bm{A}  \right)^\dagger \otimes \mathbf{I} \,.
\label{eq:inverse}
\end{equation}
{Note that $\mathbb{L}_0^\dagger$ depends only on the inter-layer coupling matrix $\bm{A}$ (e.g., for nearest-neighbor-in-time coupling, $\bm{A}=\bm{A}^\textrm{(chain)}$), which one can compute and save in memory prior to analyzing network data.}

Just as we examined first-order terms to solve for constants $\{\alpha_i\}$, we now seek a solvability condition in the second-order terms to determine $\{\beta_i\}$ in Eq.~\eqref{eq:v1}. Substituting $\lambda_\textrm{max}(\epsilon) = \lambda_0+\epsilon\lambda_1+\epsilon^2\lambda_2$ and $\mathbbm{v}(\epsilon) = \mathbbm{v}_0 + \epsilon \mathbbm{v}_1+\epsilon^2\mathbbm{v}_2$ into Eq.~\eqref{eq:eval1} and collecting the second-order terms yields
\begin{equation}
	\mathbb{L}_0\mathbbm{v}_2 = \mathbb{L}_1\mathbbm{v}_1 - \lambda_2 \mathbbm{v}_0 \,. \label{eq:second0}
\end{equation}
Similar to before, we left-multiply Eq.~\eqref{eq:second0} by $(\mathbb{P}\mathbbm{u}_i)^T$ and require both sides to be identically $0$ to obtain
\begin{equation}
	\lambda_2  \mathbbm{u}_i^T\mathbb{P}^T \mathbbm{v}_0 = \mathbbm{u}_i^T\mathbb{P}^T \mathbb{L}_1\mathbbm{v}_1 
= \mathbbm{u}_i^T\mathbb{P}^T\mathbb{G}\mathbbm{v}_1 -\lambda_1\mathbbm{u}_i^T\mathbb{P}^T\mathbbm{v}_1 .
\end{equation}
Using $\alpha_i  = \mathbbm{u}_i^T\mathbb{P}^T \mathbbm{v}_0$ and $\beta_i = \mathbbm{u}_i^T\mathbb{P}^T \mathbbm{v}_1$, it then follows that
\begin{equation}\label{here}
	\lambda_2  \alpha_i + \lambda_1\beta_i = \mathbbm{u}_i^T\mathbb{P}^T\mathbb{G}\mathbbm{v}_1\, .
\end{equation}
Substituting the expressions for $\mathbbm{v}_0$ from Eq.~\eqref{eq:v0} and $\mathbbm{v}_1$ from Eq.~\eqref{eq:v1} into Eq.~\eqref{here} then yields
\begin{equation}
	\lambda_2  \alpha_i + \lambda_1\beta_i 
=  \sum_j \alpha_j \mathbbm{u}_i^T\mathbb{P}^T\mathbb{G} \mathbb{L}_0^{\dagger} \mathbb{G} \mathbb{P}\mathbbm{u}_j + \sum_j \beta_j  \mathbbm{u}_i^T\mathbb{P}^T\mathbb{G}\mathbb{P}\mathbbm{u}_j \, . \label{eq:betas1}
\end{equation}
After some rearranging, we obtain
\begin{align}
	(\mathbf{X}^{(1)} -\lambda_1\mathbf{I}) \bm{\beta} &= (\lambda_2\mathbf{I} -  \mathbf{X}^{(2)}) \bm{\alpha}  \,,
\label{eq:betas2}
\end{align}
where the matrix $\mathbf{X}^{(1)}$ was defined by Eq.~\eqref{eq:M1star_0}, and the elements of the matrix $\mathbf{X}^{(2)}$ are
\begin{equation}
	X^{(2)}_{ij} = \mathbbm{u}_i^T\mathbb{P}^T \mathbb{G} \mathbb{L}_0^\dagger \mathbb{G}  \mathbb{P}\mathbbm{u}_j \,.\label{eq:X2}
\end{equation} 
Recalling that we determined $\bm{\alpha}$ as the solution of $\mathbf{X}^{(1)} \bm{\alpha} =  \lambda_1 \bm{\alpha}$ such that $\sum_i {\alpha}^2_i=1$, we left-multiply Eq.~\eqref{eq:betas2} by $\bm{\alpha}^T$ to obtain
\begin{equation}
	\lambda_2 =  \bm{\alpha}^T \mathbf{X}^{(2)} \bm{\alpha} \label{eq:lam2}\,. 
\end{equation}
\drt{We thereby obtain}
\begin{align}
	\bm{\beta} &=(\mathbf{X}^{(1)} -\lambda_1\mathbf{I})^\dagger(\lambda_2\mathbf{I} -  \mathbf{X}^{(2)}) \bm{\alpha} + b\bm{\alpha}\,, \label{eq:betas3}
\end{align}
where the constant $b = \bm{\alpha}^T \bm{\beta}$ describes the (possibly nonzero) projection of $\bm{\beta}$ onto the null space of $(\mathbf{X}^{(1)} -\lambda_1\mathbf{I})$ [see Eq.~\eqref{eq:ap1}]. 

We now show that $b=0$ in Eq.~\eqref{eq:betas3}, by virtue of the requirement that the eigenvector obtained at first-order has a norm of $1$. That is, we require that
$1 = \| \mathbbm{v}_0 + \epsilon \mathbbm{v}_1 \|^2
= \| \mathbbm{v}_0\|^2 +  2\epsilon \langle\mathbbm{v}_0,\mathbbm{v}_1\rangle + O(\epsilon^2)$. 
However, $\| \mathbbm{v}_0\|^2 = 1$, so $\langle\mathbbm{v}_0,\mathbbm{v}_1\rangle=0$, where we use the notation $\langle \cdot,\cdot \rangle$ to denote the dot product between inputs. Using the definitions of $\mathbbm{v}_0$ and $\mathbbm{v}_1$, we see that
\begin{align}
	0 = \left\langle \mathbbm{v}_0,\mathbbm{v}_1\right\rangle 
&= \left\langle \sum_j \alpha_j \mathbb{P}\mathbbm{u}_j, \,\, \mathbb{L}_0^{\dagger}\mathbb{G}\mathbbm{v}_0 + \sum_j \beta_j \mathbb{P}\mathbbm{u}_j\right\rangle \nonumber\\
&= \sum_j \alpha_j\beta_j 
	= b\,,
\end{align}
because the vectors $\{\mathbb{P}\mathbbm{u}_j\}$ are orthonormal and lie in the null space of $\mathbb{L}_0$. (In other words, $(\mathbb{P}\mathbbm{u}_i)^T \mathbb{P}\mathbbm{u}_j = \delta_{ij}$ and $(\mathbb{P}\mathbbm{u}_i)^T\mathbb{L}_0^{\dagger}=0$.)

In practice, we solve Eq.~\eqref{eq:betas3} using a linear solver (see, e.g., \cite{golub}) rather than the pseudoinverse to avoid computing the inverse of $(\mathbf{X}^{(1)} -\lambda_1\mathbf{I})$. We then ensure that the solution is orthogonal to $\bm{\alpha}$ {by projecting it onto the subspace that is orthogonal to $\bm{\alpha}$}.

One can substitute the solution $\bm\beta$ to Eq.~\eqref{eq:betas3} with $b=0$ into Eq.~\eqref{eq:v1} to obtain the first-order term $\mathbbm{v}_1$ in the expansion for $\mathbbm{v}(\epsilon)$. This first-order term, which yields the strongest temporal variation of the conditional centralities at small $\epsilon$, is \drt{a} concise representation of temporal changes in centrality. \drt{There are multiple possible ways to use $\mathbbm{v}_1$ to quantify the role of physical node $i$ across the $T$ layers. We define} a measure $m_i$ that equals the square root of the sum of the squares of the entries in $\mathbbm{v}_1$ \drt{that correspond} to physical node $i$. 
Specifically, we define the \emph{first-order-mover score} $m_i \ge0$ of physical node $i$ by 
\begin{align}
	m^2_i &= \mathbbm{v}_1^T \mathbb{P} \mathbbm{I}_i \mathbb{P}^T \mathbbm{v}_1 \nonumber \\
&=  \beta_i^2 + \left(\mathbb{L}_0^{\dagger}\mathbb{G}\mathbbm{v}_0\right)^T \mathbb{P} \mathbbm{I}_i \mathbb{P}^T\mathbb{L}_0^{\dagger}\mathbb{G}\mathbbm{v}_0   \nonumber\\
	&=  \beta_i^2 + \sum_{t=1}^T \left([\mathbb{L}_0^{\dagger}\mathbb{G}\mathbbm{v}_0]_{i+t(N-1)}\right)^2\,, \label{eq:fmover}
\end{align}
where $[\cdot]_i$ denotes the $i$th entry in a vector and $\mathbbm{I}_i={\diag}[0,\dots,0,\bm{I},0,\dots,0]$ is a matrix of size $NT\times NT$ that contains all $0$ entries except for the $i$th block, which is an identity matrix $\bm{I}$ of size $T\times T$. In other words, we measure the variation of $\mathbbm{v}_1$ with respect to a physical node $i$ by examining the 2-norm of the entries in $\mathbbm{v}_1$ that correspond to the node-layer pairs $(i,t)$ that are relevant to physical node $i$ (i.e.,~entries $j$ such that $j=N(t-1)+i$, with $t=1,\dots,T$). In principle, \drt{one can also use a different vector norm or a heuristic method for aggregating centrality.} Our choice has the virtue that it is mathematically consistent with our definition for the time-averaged centralities $\{\alpha_i\}$ [see Eq.~\eqref{eq:ap1}]. Specifically, $\alpha_i^2 = \mathbbm{v}_0^T \mathbb{P} \mathbbm{I}_i \mathbb{P}^T \mathbbm{v}_0$. Therefore, one can naturally extend our approach for quantifying the contribution of the first-order correction given by Eq.~\eqref{eq:fmover} to higher-order corrections. 
{We also note that first-order-mover scores rank the nodes according to the magnitudes of their corresponding entries in $\mathbbm{v}_1$. Therefore, the associated centralities $\mathbbm{v}(\epsilon)$ can either increase or decrease over time. One can easily check whether \drt{there} is an increase or a decrease by examining the corresponding entries \drt{of} $\mathbbm{v}(\epsilon)$.}

\subsection{Procedure for Computing Time-Averaged Centrality and First-Order-Mover Scores}\label{sec:algorithm}

We summarize our procedure for computing time-averaged centrality and first-order-mover scores: 
\drt{
\begin{enumerate}
\item{Construct the matrix $\mathbf{X}^{(1)}$ using Eq.~\eqref{eq:M1star}:
\begin{align}
	{X}^{(1)}_{ij} &= \gamma_1^{-1}\sum_{t} {C}^{(t)}_{ij} \sin^2 \left(\frac{\pi t}{T+1}\right)\,. \nonumber
\end{align} 
When layers are coupled by a layer-adjacency matrix $\bm{A}$ (which is not necessarily an undirected chain $\bm{ A}^{(\textrm{chain})}$), it follows that ${X}^{(1)}_{ij} = \mathbbm{u}_i^T\mathbb{P}^T \mathbb{G}  \mathbb{P}\mathbbm{u}_j $, where $\mathbb{G} = \text{diag}[ \mathbf{C}^{(1)},\dots, \mathbf{C}^{(T)}]$ and $\mathbb{P}$ and $\mathbbm{u}_i$ are defined just before and after, respectively, Eq.~\eqref{eq:perm}.
}
\item{Solve for the time-averaged centralities $\{\alpha_i\}$ using Eq.~\eqref{eq:ap1}:
\begin{equation}
	\mathbf{X}^{(1)} \bm{\alpha} =  \lambda_1 \bm{\alpha}\,.\nonumber
\end{equation}
}
\item{Construct the matrix $\mathbf{X}^{(2)}$ using Eq.~\eqref{eq:X2}:
\begin{equation}
	X^{(2)}_{ij} = \mathbbm{u}_i^T\mathbb{P}^T \mathbb{G} \mathbb{L}_0^\dagger \mathbb{G}  \mathbb{P}\mathbbm{u}_j \,,\nonumber
\end{equation} 
where $\mathbb{L}_0^\dagger  = \left(\lambda_0\bm{I} - \bm{A}  \right)^\dagger \otimes \mathbf{I}$.}
\item{Solve for $\bm{\beta}$ in Eq.~\eqref{eq:betas2}:
\begin{align}
(\mathbf{X}^{(1)} -\lambda_1\mathbf{I}) \bm{\beta} &= (\lambda_2\mathbf{I} -  \mathbf{X}^{(2)}) \bm{\alpha}  \,,
\nonumber
\end{align}
where $\lambda_2=\bm{\alpha}^T \mathbf{X}^{(2)} \bm{\alpha}$.
}
\item{Solve for the first-order-mover scores $\{m_i\}$ using Eq.~\eqref{eq:fmover}:
\begin{align}
m^2_i &=  \beta_i^2 + \sum_{t=1}^T \left([\mathbb{L}_0^{\dagger}\mathbb{G}\mathbbm{v}_0]_{i+t(N-1)}\right)^2\,,
\nonumber 
\end{align}
where $\mathbbm{v}_0=\sum_j\alpha_j \mathbb{P}\mathbbm{u}_j$.
}
\end{enumerate}
}

\medskip
\medskip

\drt{We comment briefly on the computational costs of the above procedure.} The supra-centrality matrix [see Eq.~\eqref{eq:sup_centrality1}], whose dominant eigenvector gives the joint node-layer centralities, has size $NT \times NT$, and that can be problematic for large networks with many time layers (i.e., when $T\gg 1$). \drt{The} time-averaged node centralities are given by the solution to Eq.~\eqref{eq:ap1}, which is a dominant eigenvalue problem for a matrix of size $N\times N$. To \drt{examine} which physical nodes have centralities that change significantly \drt{over} time, we examine the first-order-mover scores given by \drt{Eq.~\eqref{eq:fmover}; this requires} one to solve the \drt{$N$-dimensional} linear system given by Eq.~\eqref{eq:betas3}. Because $\mathbbm{L}_0^\dagger$, $\mathbbm{G}$, and $\mathbbm{v}_0$ are known prior to solving Eq.~\eqref{eq:betas3}, \drt{one} can directly compute the second term in Eq.~\eqref{eq:fmover}. For sparse networks [i.e., those in which the number of edges at a given time is $\mathcal{O}(N)$], the matrices that we have discussed in this section are typically also sparse. One can thus solve Eqs.~\eqref{eq:ap1}, \eqref{eq:M1star}, \eqref{eq:betas2}, \eqref{eq:X2}, and \eqref{eq:fmover} efficiently using data structures that are designed for sparse matrices, including direct \drt{methods \cite{davis2006direct}, iterative methods \cite{saad2003iterative}, and methods designed} for particular network structures (e.g., nested dissection for planar networks \cite{lipton1979generalized}). In particular, the power method for computing a dominant eigenvalue and eigenvector of a sparse matrix reduces the \drt{per-iteration} complexity from $\mathcal{O}(N^2)$ to $\mathcal{O}(M)$, where $M$ is the number of nonzero entries in the sparse \drt{matrix. However, the actual scaling can be much larger, because the number of iterations required for convergence depends on the gap between the largest and second-largest eigenvalues.}

\section{Case Studies with Empirical Network Data}\label{sec:num}

In this section, we examine temporal centrality in case studies with three sets of empirical data: the Mathematics Genealogy Project (MGP; see Sec.~\ref{sec:mgp}) \drt{of Ph.D. receipt in the mathematical sciences in} U.S. universities, top billing in the Golden Age of Hollywood (GAH; see Sec.~\ref{sec:hollywood}), and citations of U.S. Supreme Court decisions (SCD; see Sec.~\ref{sec:supremes}). \drt{We have posted {\sc Matlab} software at \cite{code} that implements our calculations and can be used to reproduce the results of this section.}

\drt{Most of our calculations for these examples use a temporal generalization of hub and authority scores \cite{kleinberg1999}, which are particularly appropriate for directed networks (such as our three examples). We also note that hub and authority scores have been used previously to examine time-independent faculty-hiring networks \cite{fowler2007,myers2011} and Supreme Court decisions \cite{fowler2007,fowler2008}. To illustrate a comparison with another choice of centrality, we also study a temporal generalization of PageRank for the MGP network.} 

\drt{We compute hub and authority scores independently using two supra-centrality matrices, ${\bf A}^{(t)}[{\bf A}^{(t)}]^T$ and $[{\bf A}^{(t)}]^T{\bf A}^{(t)}$. For a single-layer network, it is possible to simultaneously compute hub and authority scores by studying the single system $\left[{0\atop {\bf A}^T}~{{\bf A}\atop 0}\right]$. In general, however, a supra-centrality matrix that uses this alternative formulation yields different time-dependent centralities from ones based on independent computations of hub and authority scores.}
\drt{For each centrality that we compute,  we also examine the induced ranking of nodes in which the highest-ranked nodes (i.e., those with ranks 1, 2, and so on) correspond to the largest centralities, whereas the lowest-ranked nodes correspond to the smallest centralities.}

\subsection{Doctoral Degree Exchange in the Mathematics Genealogy Project (MGP)}\label{sec:mgp}

Our first case study \drt{uses} a network that encodes the exchange of mathematicians (and other mathematical scientists) who have obtained a Ph.D. (or equivalent doctoral degree) between universities to study the academic prestige of those universities. We study data provided by the Mathematics Genealogy Project (MGP) \cite{mgp}, which collects information for mathematicians (and members of related fields who are listed in the MGP) with doctorates. For each mathematical scientist, the information includes graduation year, his/her official academic advisor(s), the degree-granting university, and a list of his/her students who have also obtained doctoral degrees. A subset of the present authors previously utilized this data to approximate the flow of doctorates between universities---that is, a person graduates from one university and is then hired at a second university---and quantified the resulting hub and authority scores for the total flow during a specified time period as a candidate measure of these universities' relative mathematical prestige \cite{myers2011}. \drt{Moreover, hub and authority scores have been used previously to study Ph.D. exchange networks in other disciplines \cite{fowler2007}.}
See \cite{han2003} for a comparison of the hiring market for different academic disciplines, and see\drt{\cite{malmgren2010,gargiulo2016classical}} for other analyses and visualization of data from the MGP. See \cite{shep2015} for an application of PageRank centrality to ranking world universities using data from Wikipedia.

It is well-documented that graduates typically obtain faculty positions at universities that are either comparable to or less prestigious than \drt{the one} from which they graduate \cite{caplow2001,burris2004,fowler2007,paneretos2012,deville2014,clauset2015}, and we study university prestige as indicated by the exchange \drt{between universities} of mathematicians with doctoral degrees. 
We generalize a previous study of the MGP data in \cite{myers2011} by keeping the year that each faculty member graduated with his/her Ph.D. degree. We focus on the years 1946--2010, which includes all post-World War II information available in the data set.\footnote{The data set was provided to us in 2009, although it includes information up to 2010. The year 2006 is the last year in which a Ph.D. degree was awarded to someone who was subsequently a Ph.D. advisor in the data, so it is also the last year in which intra-layers edges are present. Additionally, we decided to be optimistic and include Ph.D. degrees that were projected for the year 2010.} This yields $T=65$ time layers, and we restrict our attention to a set of $N=227$ U.S. universities that were connected during this period. To construct the network, we create directed intra-layer edges $i\to j$ at time $t$ to represent a doctoral degree in the MGP data \drt{awarded to a mathematical scientist from} university $j$ in year $t$ who later advised at least one student at university $i$. \drt{Therefore}, to contribute a directed edge, the \drt{mathematician} must have at least one student in the MGP data. We weight edges to indicate the number of doctorates from university $j$ in year $t$ \drt{awarded to faculty} who later advise students at university $i$. Our construction aligns edge directions to be opposite to that of the flow of people, so a node with large in-degree (i.e., with many graduates who \drt{later} advise students elsewhere) is considered both an academic authority as well as an authority with respect to HITS centrality \cite{kleinberg1999}. See Fig.~\ref{fig:MGP_network} for a visualization of this network; due to the difficulty of visualizing temporal networks \drt{\cite{de2014muxviz}}, we \drt{show} a network corresponding to the aggregation $\sum_t \mathbf{A}^{(t)}$ of the adjacency matrices. Although \drt{one} could define the multilayer network in more intricate ways (e.g., by normalizing edge weights using the number of graduates) and examine how the results vary for different choices, we wish to keep the present manuscript focused on introducing and demonstrating our temporal generalization of eigenvector-based centralities. Therefore, we leave such detailed analyses for future work. \drt{We make the MGP temporal network available as a Supplementary Material and online at \cite{MGP_data}.}

\begin{table}[t!]
\caption{Top time-averaged centralities and first-order-mover scores for our temporal generalization [see Eq.~\eqref{eq:sup_centrality1}] of authority scores \cite{kleinberg1999} for U.S. universities in the Mathematics Genealogy Project \cite{mgp}. 
\drt{We show our results using two different orderings: (1) according to the top time-averaged centralities and (2) according to the top first-mover scores.}
}
\centering 
\small{
\begin{tabular}{c c c} 
~&Top Time-Averaged Centralities & Top First-Order-Mover Scores \\
\begin{tabular}{c } 
\hline\hline 
Rank  \\ [0.5ex] 
\hline 
1  \\ 
2  \\
3  \\
4   \\
5  \\ 
6  \\ 
7   \\ 
8   \\ 
9   \\ 
10   \\ 
[1ex] 
\hline 
\end{tabular}
&
\begin{tabular}{c c c} 
\hline\hline 
 University & $\alpha_i$& $m_i$  \\ [0.5ex] 
\hline 
MIT & 0.6685  & \drt{688.62} \\ 
UC Berkeley & 0.2722 & \drt{299.06}\\
Stanford & 0.2295 &  \drt{241.71}\\
Princeton & 0.1803 &\drt{248.71} \\
UIUC & 0.1645  &\drt{74.30} \\ 
Cornell & 0.1642 & \drt{180.50} \\ 
Harvard & 0.1628  &\drt{185.34} \\ 
U Washington & 0.1590 &\drt{81.22}  \\ 
U Michigan & 0.1521 &\drt{86.50} \\ 
UCLA & 0.1456  &\drt{152.77} \\ 
[1ex] 
\hline 
\end{tabular}
&
\begin{tabular}{c c c} 
\hline\hline 
University & $m_i$  & $\alpha_i $\\ [0.5ex] 
\hline 
MIT & 688.62 &\drt{0.6685}\\ 
UC Berkeley & 299.07&\drt{0.2722} \\
Princeton & 248.72  &\drt{0.1803} \\
Stanford & 241.71&\drt{0.2295}  \\
Georgia Tech & 189.34 &\drt{0.0960}  \\ 
U Maryland & 186.65  &\drt{0.1278} \\ 
Harvard & 185.34&\drt{0.1628}   \\ 
CUNY & 182.59 &\drt{0.0466}  \\ 
Cornell & 180.50&\drt{0.1642}  \\ 
Yale & 159.11  &\drt{0.0816} \\ 
[1ex] 
\hline 
\end{tabular}
\end{tabular}\\
\label{table:MGP}
}
\end{table}

\subsubsection{{MGP: Centrality in the Strong-Coupling Regime}}\label{sec:mgp_singular}

We begin by identifying the universities that have the largest time-averaged authority centralities $\{\alpha_i\}$, which we obtain from the \drt{dominant} eigenvector of the matrix $\mathbf{X}^{(1)}$ [see Eq.~\eqref{eq:M1star}] using $\mathbf{C}^{(t)}=(\mathbf{A}^{(t)})^T\mathbf{A}^{(t)}$ \cite{kleinberg1999}. 
For notational convenience, we use $t$ to denote the graduation year rather than the time layer. For example, we use $\mathbf{A}^{(1946)}$ to denote the network adjacency matrix for time layer 1 (i.e., year 1946).
\drt{We summarize these authority values in Table~\ref{table:MGP}}, and we note that the most central universities according to this measure are all widely-accepted top-tier programs in mathematics. The time-averaged authorities identify the four most central mathematics universities for this time period as MIT, UC Berkeley, Stanford, and Princeton. Although the results in Table \ref{table:MGP} are interesting, time-averaged centrality (by definition) does not provide information about temporal trajectories of the universities' authorities, and this is the type of idea that we seek to explore. We thus calculate the first-order-mover scores $\{m_i\}$ from Eq.~\eqref{eq:fmover}, and we list the universities with \drt{the} top first-order-mover scores in the right column of Table \ref{table:MGP}. Note the similarity between the two lists; that is, universities with \drt{largest $\alpha_i$} tend to also have \drt{large $m_i$}.

In Fig.~\ref{fig:MGP1}, we show further results for prestige (as revealed by Ph.D. exchange). 
In Fig.~\ref{fig:MGP1}(a), we plot university ranking according to $\{m_i\}$ versus its ranking according to $\{\alpha_i\}$. Note in the bottom left corner that MIT is ranked first for both quantities, and that in general there is a strong linear correlation between rank according to $\alpha_i$ and rank according to $m_i$. Intuitively, this suggests that shifts in centrality include a natural effect that is related directly to the centrality score itself. (In other words, large centrality values tend to also include large fluctuations, whereas small centrality values typically \drt{have} only small fluctuations.) Deviations from the observed nearly-linear relation indicate universities whose centrality trajectory exhibits larger variations over time, and it is worthwhile to look at these universities in more detail for potentially interesting insights. For example, the universities with \drt{high rank according to $m_i$  (i.e., large $m_i$)} but comparatively low rank according to $\alpha_i$ (i.e., small $\alpha_i$)} include Georgia Tech and CUNY, and it is known that Georgia Tech's mathematics department transitioned from a primarily teaching-oriented department to a much more research-oriented department with a newly restructured doctoral degree program starting in the late 1970s \cite{DukeGT}. 

In Fig.~\ref{fig:MGP1}(b), we plot the conditional authority centralities at $\epsilon = 10^{-4}$ of universities versus time for six of the universities with the largest first-order-mover scores $m_i$. This includes the four universities with the top time-averaged centralities, as well as Georgia Tech and CUNY (which do not have highly-ranked time-averaged centralities). As we expect, the conditional centralities for Georgia Tech and CUNY change drastically over time, whereas the trajectories for the others remain relatively constant.

\begin{figure}[t!]
\centering
\includegraphics[width=.495\linewidth]{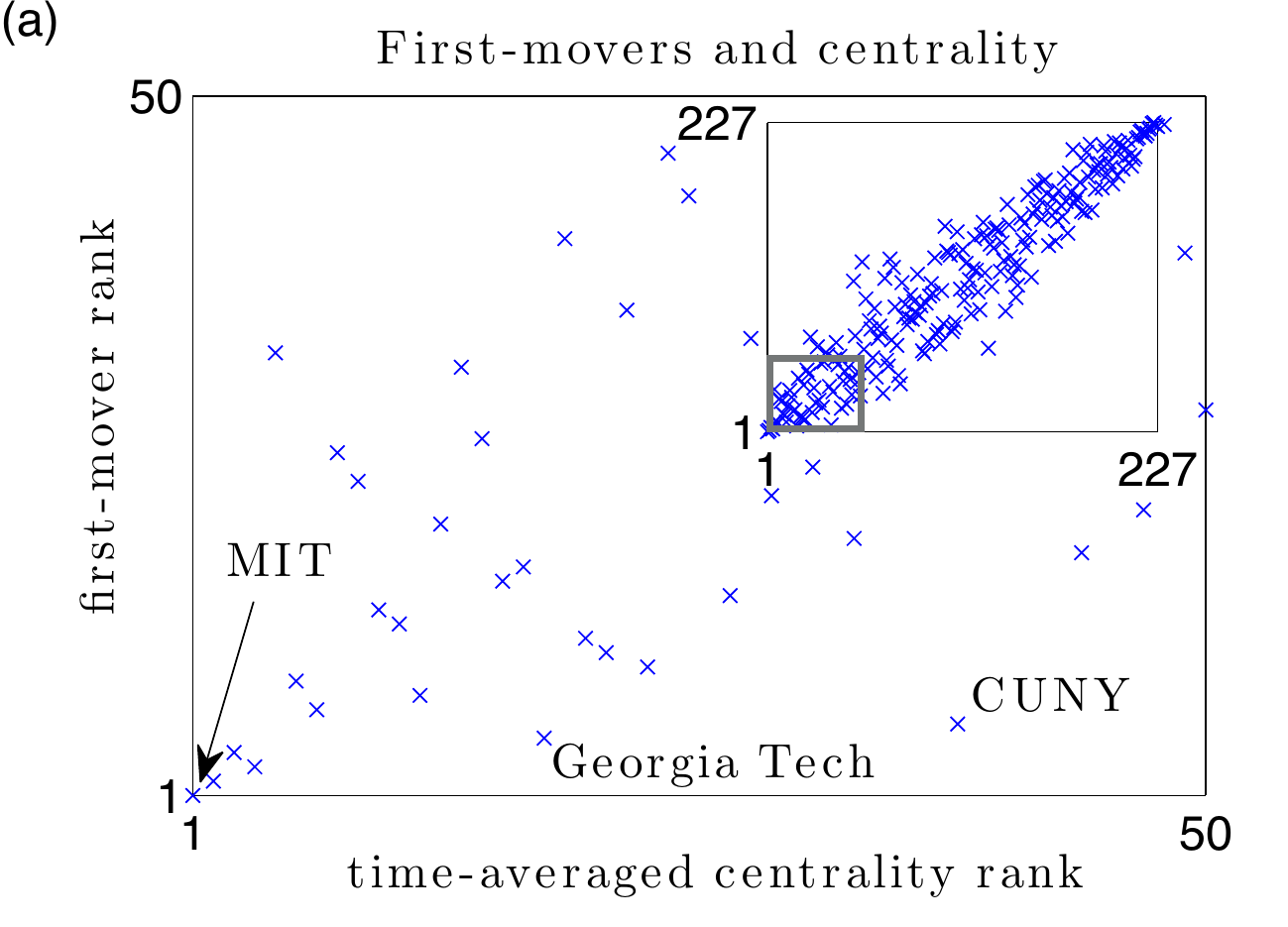}
\includegraphics[width=.495\linewidth]{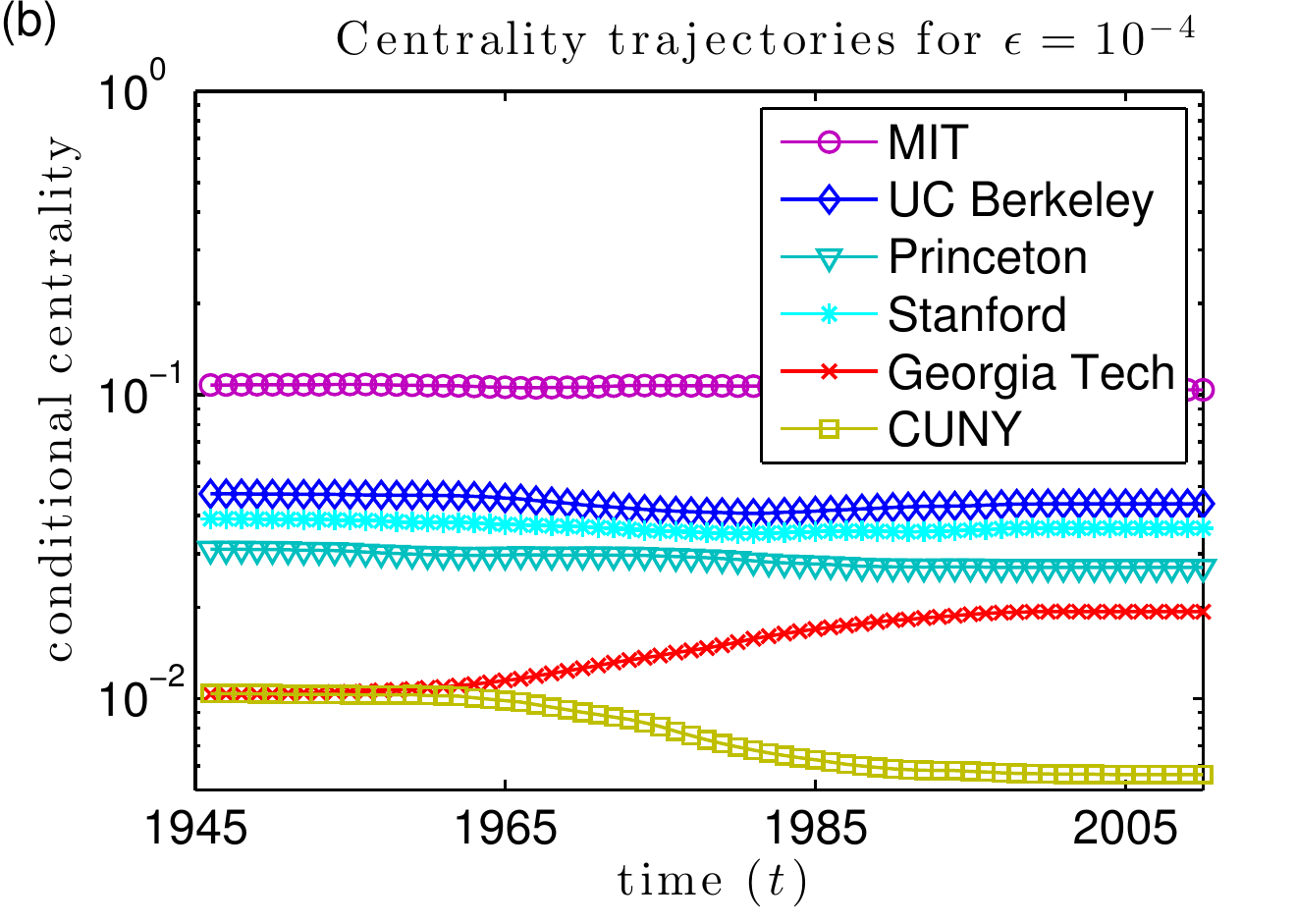}
\caption{
University rankings in the Mathematics Genealogy Project (MGP) \cite{mgp} according to time-averaged centralities and first-order-mover scores. We give results for our temporal generalization [see Eq.~\eqref{eq:sup_centrality1}] of authority scores \cite{kleinberg1999}.
(a) We plot the first-order-mover ranking of nodes (i.e., ranked according to $\{m_i\}$) versus the time-averaged ranking of nodes (i.e., ranked according to $\{\alpha_i\}$).  As shown in the inset, nodes with large time-averaged rank tend to also have large first-order-mover rank (e.g., MIT ranks first in both). However, there are nodes that have a much higher first-order-mover rank than time-averaged \drt{centrality} rank (e.g., Georgia Tech and CUNY). In panel (a), we show a magnification of the gray box \drt{in} the subpanel.
(b) We plot the conditional node-layer centralities (i.e., the centrality of node-layer pairs normalized by the centrality of each time layer) to study the universities' centrality trajectories over time. We show results for some of the top ranked first-order-movers. Most of these top first-order-movers are also top time-averaged authorities (e.g., MIT). In contrast, Georgia Tech and CUNY rank in the top six of the first-order-movers ranking, but they are in the lower reaches of the top 40 for the time-averaged ranking. As expected, this ranking difference reflects the fact that their centrality trajectories exhibit a significant change over time. Georgia Tech rises in rank when $t\in[1965,1985]$, whereas CUNY's rank drops during this time period.
}
\label{fig:MGP1}
\end{figure}

\subsubsection{{MGP: Some Properties of \drt{Authority} Centrality}}\label{sec:mgp_props}

As we showed in Sec.~\ref{sec:toy} for a synthetic network, the choice of inter-layer coupling strength $\epsilon$ strongly affects the temporal behavior of a node's centrality trajectory. We expect to observe three qualitative regimes: (i) the strong-coupling ($\epsilon\to0^+$) regime that we studied in Sec.~\ref{sec:TimeAve}; (ii) a weak-coupling ($\epsilon\to\infty$) regime; and (iii) an intermediate-coupling regime, in which the centralities behave differently than expected for either the \drt{strong-coupling} or weak-coupling regimes. In Fig.~\ref{fig:MGP1}(b), we show results for $\epsilon=10^{-4}$, and we observe that the universities tend to have slowly-varying centrality trajectories. However, the choice of $\epsilon$ should depend both on the application and on the question of interest. As we are about to illustrate, it is important to consider what values of $\epsilon$ are appropriate.

\begin{figure}[t!]
\centering
\includegraphics[width=.495\linewidth]{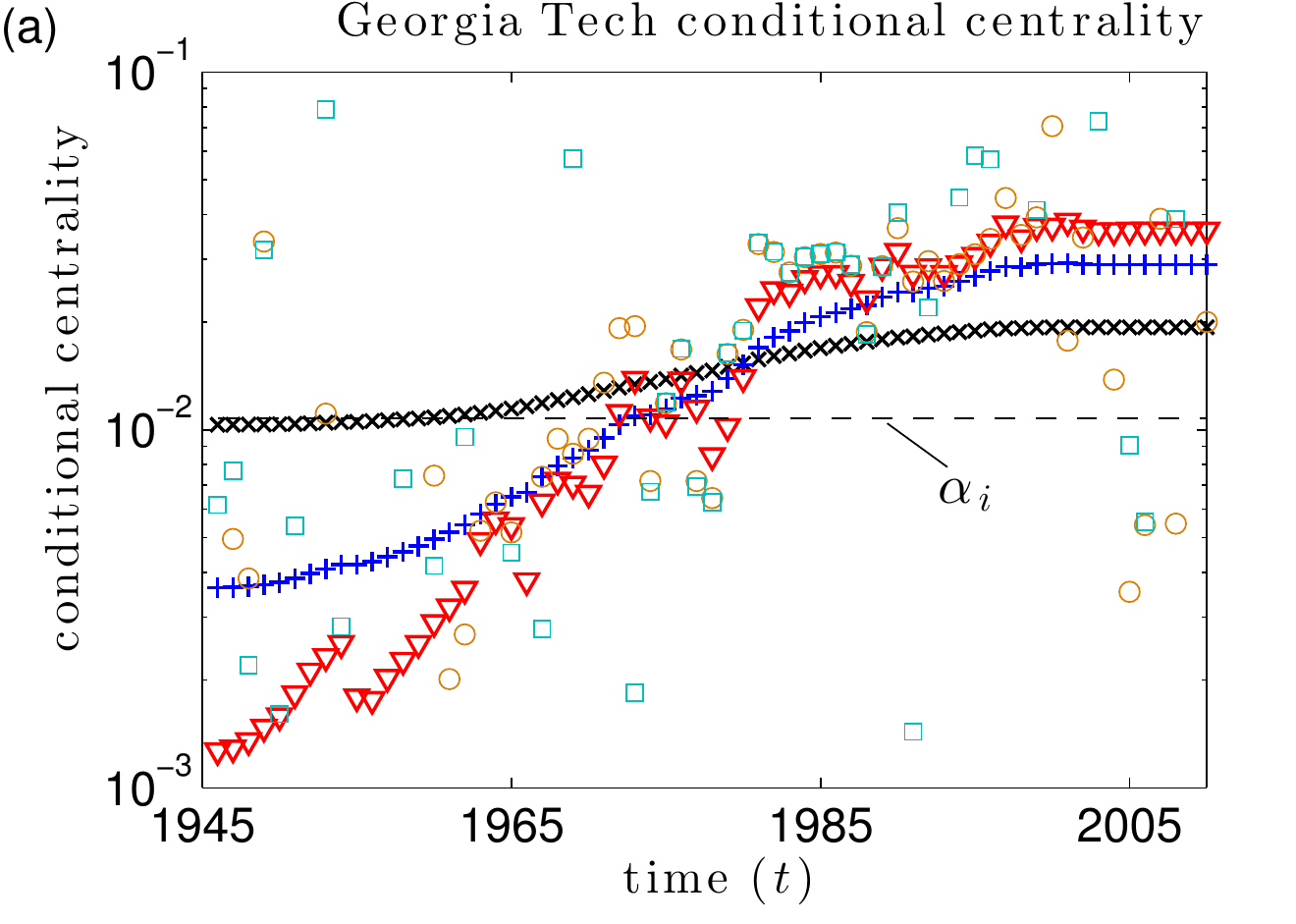}
\includegraphics[width=.495\linewidth]{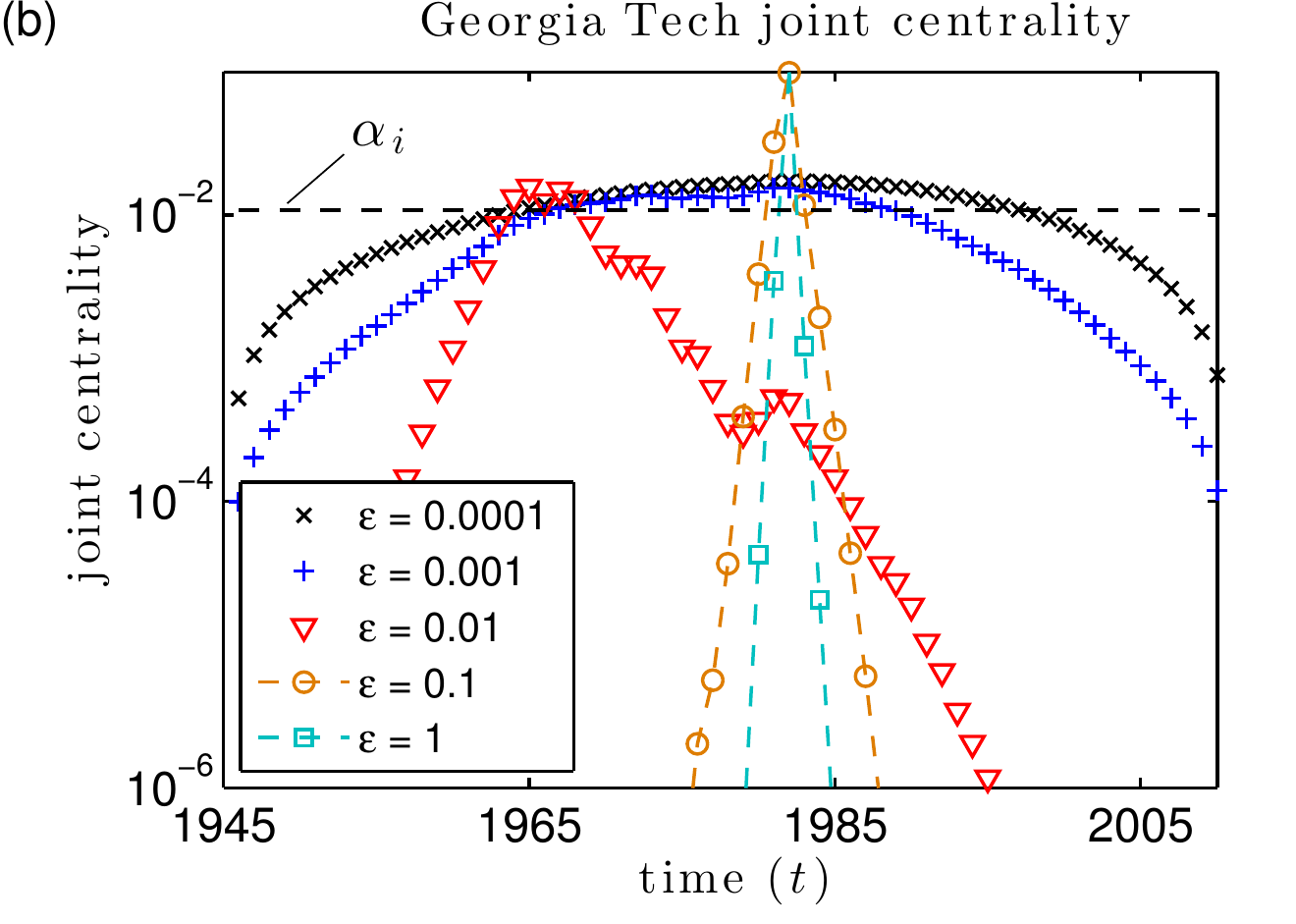}
\caption{  
Centrality trajectories for Georgia Tech illustrate that one can construe $\epsilon$ as a tuning parameter that controls how much centrality can vary between neighboring time layers. 
(a)~To study the trajectory of university authorities over time, we examine the conditional node-layer centralities. For sufficiently small $\epsilon$, we observe a steady increase in ranking with time for Georgia Tech. Varying $\epsilon$ changes the coupling strength between temporal layers. If $\epsilon$ is too large (e.g., $\epsilon\ge0.01$), then the coupling between layers is so weak that the conditional node-layer centrality of the two node-layer pairs at times $t$ and $t+1$ are no longer similar in value. As $\epsilon\to0^+$, the conditional node-layer centrality \drt{approaches} the stationary, time-average ranking given by $\alpha_i$ (horizontal dashed line), but we still observe significant variation even for $\epsilon=0.0001$.
(b) We plot the joint node-layer centrality of the $T$ node-layer pairs that correspond to Georgia Tech across the time layers for several values of $\epsilon$. For small $\epsilon$, the joint node-layer centralities are determined by the chain of identity edges (which leads to the sinusoidal dependence given by Eq.~\eqref{eq:chain_2} with $n=1$ and magnitude $\alpha_i$). For large values of $\epsilon$, the node-layer pairs in time layer $t=1982$ dominate the joint node-layer centralities, so all universities have their highest values in this layer. 
}
\label{fig:GT1}
\end{figure}

In Fig.~\ref{fig:GT1}, we study centrality trajectories for Georgia Tech \drt{for} various choices for $\epsilon$. In Fig.~\ref{fig:GT1}(a), we show the conditional node-layer centralities for Georgia Tech versus time $t$. Recall that the conditional node-layer centrality indicates the centrality of node-layer pair $(i,t)$ with respect to all node-layer pairs $(j,t)$ at time $t$.
We also show the value of $\alpha_i$ (rescaled for normalization), which gives the conditional node-layer centrality of Georgia Tech in the limit $\epsilon\to0^+$. For small but nonzero $\epsilon$ (e.g., $\epsilon=10^{-3}$), note that we obtain a similar trajectory as for $\epsilon=0^+$. For example, the trajectory varies slowly over time, so the conditional node-layer centrality of Georgia Tech at times $t$ and $t+1$ are approximately equal for all $t$. However, as we increase $\epsilon$, we lose the slow temporal variation over time. 
For example, when $\epsilon\ge10^{-1}$, the conditional centrality of Georgia Tech at times $t$ and $t+1$ are typically very dissimilar, which appears to be a consistent property of conditional node-layer centralities in the limit $\epsilon\to\infty$. 
It is our believe that the highly volatile rankings for large $\epsilon$ do not appropriately describe the dynamics of department prestige \cite{sorz2015}; this observation has motivated us to focus on the small $\epsilon$ (i.e., strong coupling) regime in this paper.
The limiting cases $\epsilon\to0^+$ and $\epsilon\to\infty$, respectively, do a good job of describing regimes with very small and very large $\epsilon$, but the intermediate (``transitional'') regime between these extremes is not straightforward to interpret. Even the boundaries between the two extreme qualitative regimes are not clear and are open to interpretation. 

In Fig.~\ref{fig:GT1}(b), we plot the joint node-layer centralities for Georgia Tech for various values of $\epsilon$. Recall that the joint node-layer centrality of the node-layer pair $(i,t)$ reflects information about both the physical node $i$ and the time layer $t$.  In the $\epsilon\to0^+$ limit, the joint node-layer centrality trajectory is given by $\alpha_i\bm{u}^{(\textrm{chain})}$, which we show using a dashed line. Interestingly, for the $\epsilon$ values that exhibit slowly-varying conditional-node-layer-centrality trajectories in panel (a) (i.e., $\epsilon\le10^{-3}$), we find that the joint node-layer centralities have a similar order of magnitude across the $T$ time layers.  For example, when $\epsilon=10^{-4}$, the joint node-layer centrality of Georgia Tech at time $t=1945$ is roughly one tenth that of Georgia Tech at time $t=1965$. In contrast, when $\epsilon\ge10^{-1}$, the joint node-layer centralities for Georgia Tech are concentrated at just a few time layers near $t=1982$. Note that the dominant eigenvalue of the centrality matrix $\mathbf{C}^{(1982)}$ for this time layer is larger than those for the other time layers. Thus, we \drt{find that} the joint node-layer centralities to localize at layer $t=1982$ as $\epsilon\to\infty$, \drt{a phenomenon that can be described using Perron--Frobenius theory \cite{meyer}}. This localization \drt{implies} that a single time layer is dominating the joint node-layer centralities, which we confirm with the observation that the marginal layer centralities are also localized at layers near $t=1982$ (not shown).

\begin{table}[t!]
\caption{
\drt{Top time-averaged centralities ($\alpha_i$) and first-order-mover scores ($m_i$) for our temporal generalization [see Eq.~\eqref{eq:sup_centrality1}] of PageRank  with $p=0.85$ for universities in the Ph.D. exchange network.} }
\centering 
\small{
\drt{ 
\begin{tabular}{c c c} 
~&Top Time-Averaged Centralities & Top First-Order-Mover Scores \\
\begin{tabular}{c } 
\hline\hline 
Rank  \\ [0.5ex] 
\hline 
1  \\ 
2  \\
3  \\
4   \\
5  \\ 
6  \\ 
7   \\ 
8   \\ 
9   \\ 
10   \\ 
[1ex] 
\hline 
\end{tabular}
&
\begin{tabular}{c c c} 
\hline\hline 
University & $\alpha_i$& $m_i$  \\ [0.5ex] 
\hline 
MIT & 0.484  &  {1.82} \\ 
Stanford & 0.402 &  {7.58}\\
UC Berkeley & 0.390 &   {10.76}\\
Harvard & 0.363 & {7.99} \\
Princeton & 0.303  & {15.01} \\ 
Cornell & 0.142 &  {2.11} \\ 
U Chicago & 0.122  & {4.28} \\ 
Yale & 0.112 & {2.36}  \\ 
Wisconsin-Madison & 0.111 & {2.09} \\ 
NYU & 0.102  & {3.28} \\ 
[1ex] 
\hline 
\end{tabular}
&
\begin{tabular}{c c c} 
\hline\hline 
University & $m_i$  & $\alpha_i $\\ [0.5ex] 
\hline 
Princeton & 15.01 & {0.303}\\ 
UC Berkeley & 10.76 & {0.390} \\
Harvard & 7.99  & {0.363} \\
Stanford & 7.58& {0.402}  \\
U Michigan & 5.17 & {0.101}  \\ 
Columbia & 5.03  & {0.088} \\ 
U Chicago & 4.28 & {0.122}   \\ 
NYU & 3.28 & {0.102}  \\ 
Carnegie Melon & 2.83 & {0.088}  \\ 
U Washington & 2.67  & {0.053} \\ 
[1ex] 
\hline 
\end{tabular}
\end{tabular}}
}
\label{table:MGP_pagerank} 
\end{table}

\subsubsection{MGP: PageRank Centrality} \label{sec:mgp_pagerank}

\drt{We also examine temporal centralities in the MGP network using centrality matrices given by PageRank matrices
\begin{equation}
{\bf P}^{(t)} = 
p  {\bf A}^{(t)}\text{diag}[d_1^{(t)},\dots,d_N^{(t)}]^{-1} + {(1-p)}  {\bf v}{\bf 1}^T ,
\label{eq:pagerank}
\end{equation}
where $d_j^{(t)}=\sum_i A_{ij}^{(t)}$ is the out-degree of node $j$, the quantity $1-p\in[0,1]$ is the so-called \emph{teleportation parameter} (also called the damping coefficient), ${\bf 1}$ is a vector of ones, and ${\bf v}$ is the personalized PageRank vector (which we set to be  ${\bf  v}=N^{-1}{ \bf 1}$). We handle \emph{dangling nodes} (i.e., nodes with out-degree $0$) by adding a single self-edge for each of these nodes. (This process of adding self-edges is sometimes called \emph{sink preferential PageRank} \cite{gleich2014}.)  By adding these edges, $ {\bf A}^{(t)}\text{diag}[d_1^{(t)},\dots,d_N^{(t)}]^{-1} $ is guaranteed to be column stochastic, so its spectral radius is $1$. PageRank centrality is equal to the right dominant eigenvector of the PageRank matrix. In our usage, ${\bf C}^{(t)}$ in Eq.~\eqref{eq:sup_centrality1} is given by ${\bf P}^{(t)}$, and the temporal extension of PageRank is the right dominant eigenvector of the corresponding supra-centrality matrix $\mathbb{{C}}(\epsilon)$.

In Table \ref{table:MGP_pagerank}, we show the universities with top time-averaged PageRank centrality and first-order-mover scores in the MGP \drt{Ph.D. exchange} network. We identify similar top universities with PageRank as we did with authority scores (see Table \ref{table:MGP}). For example, both tables identify MIT as the university with the top time-averaged centrality. Interestingly, the gap between $\alpha_i$ for MIT and the second-ranked university is much smaller for PageRank than it is for authority scores. We hypothesize that one major factor that contributes to this difference is the impact of self-edges. In particular, we identify 172 self-edges for MIT during the period 1946--2010, whereas the second-largest number of self-edges is only 59 (for both Stanford and UC Berkeley). This suggests that it may be interesting to examine variants of PageRank that use a different strategy to deal with dangling nodes.
}

\subsubsection{{MGP: Summary}}\label{sec:mgp_summary}

Our case study illustrates practical considerations and techniques that are useful for understanding centrality in temporal (and other multilayer) networks. \drt{In our analysis of authority centralities,} we identified two important characteristics of centrality trajectories: ``slow variation'' and ``layer localization'', which we observe, respectively, in the limits $\epsilon\to0^+$ and $\epsilon\to\infty$.  As we have demonstrated, a qualitative comparison of centrality trajectories to these limiting cases is helpful for quantifying regimes of large and small $\epsilon$ (e.g., $\epsilon\ge10^{-1}$ and $\epsilon\le10^{-3}$ for Fig.~\ref{fig:GT1}). In this example, we found for sufficiently small $\epsilon$ that the centrality trajectories \drt{vary slowly} with time. As $\epsilon$ increases, we found a \drt{transition, which we observed} in two ways: the joint centrality localizes to just a few layers, and the conditional centrality begins to exhibit large fluctuations from one time layer to the next (i.e., trajectories no longer \drt{vary slowly}). We believe this weak-coupling regime to be inappropriate for the MGP \drt{Ph.D. exchange network}, as mathematics department prestige should not fluctuate wildly from one year to the next \cite{sorz2015}. Instead, it should change on a slower time scale. The strong-coupling regime is described by our singular perturbation ($\epsilon\to0^+$) analysis. The transition between the weak-coupling and strong-coupling regimes can be rather complicated. For example, see our calculations for Georgia Tech in Fig.~\ref{fig:GT1}. Obtaining a complete description of the dependency on $\epsilon$ of centrality trajectories for all universities (i.e., not just Georgia Tech) is even more complicated. For scenarios in which exploring various $\epsilon$ is not computationally feasible, we highlight that restricting one's attention to the limit $\epsilon\to0^+$ can still yield very informative results (e.g., see Fig.~\ref{fig:MGP1}), and obviously it is much more \drt{efficient computationally}.

We have also observed fascinating phenomena, and such phenomena and our techniques for investigating them provide avenues for further study. One \drt{phenomenon} is eigenvector localization for multilayer networks.
Eigenvector localization \drt{is well-known and has received considerable attention for time-independent, single-layer networks} \cite{pastor2015distinct,kawamoto2015localized,hashimoto2015,cucuringu2011,restrepo2004,mitrovic2009,dorogovtsev2003,farkas2001,goh2001,chauhan2009,monasson1999}. Localization can arise from various forms of structural heterogeneity---including the presence of \drt{high-degree} nodes, community structure, clustering, core--periphery structure, and edge weighting. For the purpose of ranking nodes with an eigenvector-based centrality, localization can sometimes be problematic, because the centrality concentrates onto a (potentially very small) subset of the nodes, and this makes it difficult to reliably rank nodes outside of that subset. This has prompted the introduction and investigation of new centrality measures that, for example, do not exhibit localization (or at least exhibit less severe localization) due to the presence of nodes with large degree \cite{kawamoto2015localized,hashimoto2015}, although localization can still arise due to other network structures \cite{pastor2015distinct}. 
We also remark that the ``non-backtracking centrality'' (also called ``Hashimoto centrality'') introduced in Ref.~\cite{hashimoto2015} is based on an eigenvector, and the framework of the present paper thus allows us to generalize it for temporal networks.

For our numerical experiments in Fig.~\ref{fig:GT1}, in the limit of large $\epsilon$, we observe localization of the eigenvector $\mathbbm{v}(\epsilon)$ onto layer $t=1982$, which is the layer whose centrality matrix $\mathbf{C}^{(1982)}$ has the largest \drt{spectral radius}. 
\drt{Given that the layers decouple in the limit $\epsilon\to\infty$, and $\mathbb{M}(\epsilon)$ thus becomes block diagonal in that limit, localization to layer $t=1982$ is well-described by Perron-Frobenius theory. Interestingly, we find for moderate $\epsilon$ (e.g., $\epsilon=0.01$) that localization appears to first occur for layer 1967, which also corresponds to a centrality matrix with a large spectral radius. Therefore, localization in this context depends both on a supra-centrality matrix's block-diagonal structure (e.g., similar to localization with community structure \cite{chauhan2009}) and on the presence of nodes with high degree (which contribute to large eigenvalues in the centrality matrices $\{\mathbf{C}^{(t)}\}$ of the layers \cite{taylor2011,taylor2012,restrepo2007}). These preliminary findings identify eigenvector localization in multilayer networks as an exciting direction for further study.}
%
\drt{We do not observe a similar localization phenomenon for our temporal generalization of PageRank, because the spectral radii of the PageRank centrality matrices that we couple into supra-centrality matrices are all equal to $1$, so no single layer dominates. Given this observation, we expect that it can be beneficial for certain applications to weight the layers' centrality matrices to control their relative importances.}

\begin{table}[t!]
\caption{Top time-averaged centralities \drt{($\alpha_i$)} and first-order-mover scores \drt{($m_i$)} for actors during the Golden Age of Hollywood (GAH). We give results for our temporal generalization [see Eq.~\eqref{eq:sup_centrality1}] of authority scores \cite{kleinberg1999}.
}
\small{
\centering 
\begin{tabular}{c c c} 
~&Top Time-Averaged Centralities & Top First-Order-Mover Scores  \\
\begin{tabular}{c} 
\hline\hline 
Rank  \\ [0.5ex] 
\hline 
1   \\ 
2  \\
3  \\
4  \\
5  \\ 
6   \\ 
7    \\ 
8   \\ 
9  \\ 
10 \\ 
[1ex] 
\hline 
\end{tabular}
&
\begin{tabular}{c c c} 
\hline\hline 
Actor & $\alpha_i$ & $m_i$  \\ [0.5ex] 
\hline 
Gable, Clark & 0.3683  & \drt{136.32} \\ 
Marx, Groucho & 0.3627  &\drt{163.34}\\
Marx, Harpo & 0.2844   &\drt{112.28}\\
Garland, Judy & 0.2820 &\drt{100.28} \\
Tracy, Spencer & 0.2681 &  \drt{98.20}\\ 
Stewart, James & 0.2371  &\drt{78.78} \\ 
Crawford, Joan & 0.2369 &\drt{90.58}  \\ 
Astaire, Fred & 0.2103 & \drt{73.29} \\ 
Marx, Chico & 0.2055  &\drt{86.39}\\ 
Cagney, James & 0.1779  &\drt{69.00} \\ 
[1ex] 
\hline 
\end{tabular}
&
\begin{tabular}{c c c} 
\hline\hline 
Actor & $m_i$ & $\alpha_i$  \\ [0.5ex] 
\hline 
Marx, Groucho & 163.34 &\drt{0.3627}  \\ 
Gable, Clark & 136.32 &\drt{0.3683} \\
Marx, Harpo & 112.28   &\drt{0.2844}\\
Garland, Judy & 100.28  &\drt{0.2820}\\
Tracy, Spencer & 98.20  & \drt{0.2681}\\ 
Crawford, Joan & 90.58  & \drt{0.2369}\\ 
Marx, Chico & 86.39   &\drt{0.2055}\\ 
Stewart, James & 78.78 &\drt{0.2371}  \\ 
Astaire, Fred & 73.29  &\drt{0.2103} \\ 
Cagney, James & 69.00  & \drt{0.1779} \\ 
[1ex] 
\hline 
\end{tabular}
\end{tabular}}
\label{table:GAH} 
\end{table}

\subsection{Top Billing in the Golden Age of Hollywood (GAH)}\label{sec:hollywood}

In our second case study, we examine the centralities of actors who are known for their performances during the so-called ``Golden Age of Hollywood'' (GAH)---a time period spanning roughly 1920--1960. To focus on the most important actors of this period, we restrict our study to 55 movies stars (26 female and 29 male) identified by Wikipedia \cite{wiki-GAH} as being notable within Hollywood's Golden Age. For these individuals, we use the Internet Movie Database (IMDb) \cite{IMDb} to define a weighted, directed, temporal network in which each node represents a movie star and each edge encodes the number of times that a pair of individuals costarred in a movie during a given time window. We consider all movies involving this set of movie stars, and we bin the data by decade over the time period 1909--2009. That is, for each 10-year\footnote{To incorporate all available data in \drt{IMDb}, our first layer represents the 11-year time window 1909--1919. All subsequent layers correspond to 10-year time windows.} time window $t$, we include a directed edge (and add a unit weight to the edge) $i\to j$ (with $i\neq j$) for each instance in which the billing position of actor $j$ is equal to or higher than that of $i$ (i.e., actor $j$ appears earlier in the credits). If the relative billing position is unknown, we add unit weight to the \drt{edges} $i\leftrightarrow j$ (i.e., we include both directed edges). It would be interesting to explore the effect of different binning and network-construction strategies. \drt{We have made the temporal network data available as Supplementary Material and at \cite{hollywood_data}.} 

Because the temporal GAH network is directed, we again choose to use a supra-centrality matrix in which the authority matrices are along the diagonal blocks. In Table \ref{table:GAH}, we list the actors with the top time-averaged authority centralities and the top first-order-mover authority scores. All of these actors appeared in some of their most famous roles during the 1930s and 1940s, and many of these roles led to prestigious awards (e.g., Oscar nominations and wins). Interestingly, the ten actors actors with highest time-averaged centralities also have the highest first-order-mover scores, and we \drt{thus} do not identify any major shifts in the centrality trajectories of these top actors over time. During this time period, men were nearly always billed higher than women, so it is not surprising that only two women appear in Table \ref{table:GAH}.

\begin{figure}[t!]
\centering
\includegraphics[width=\linewidth]{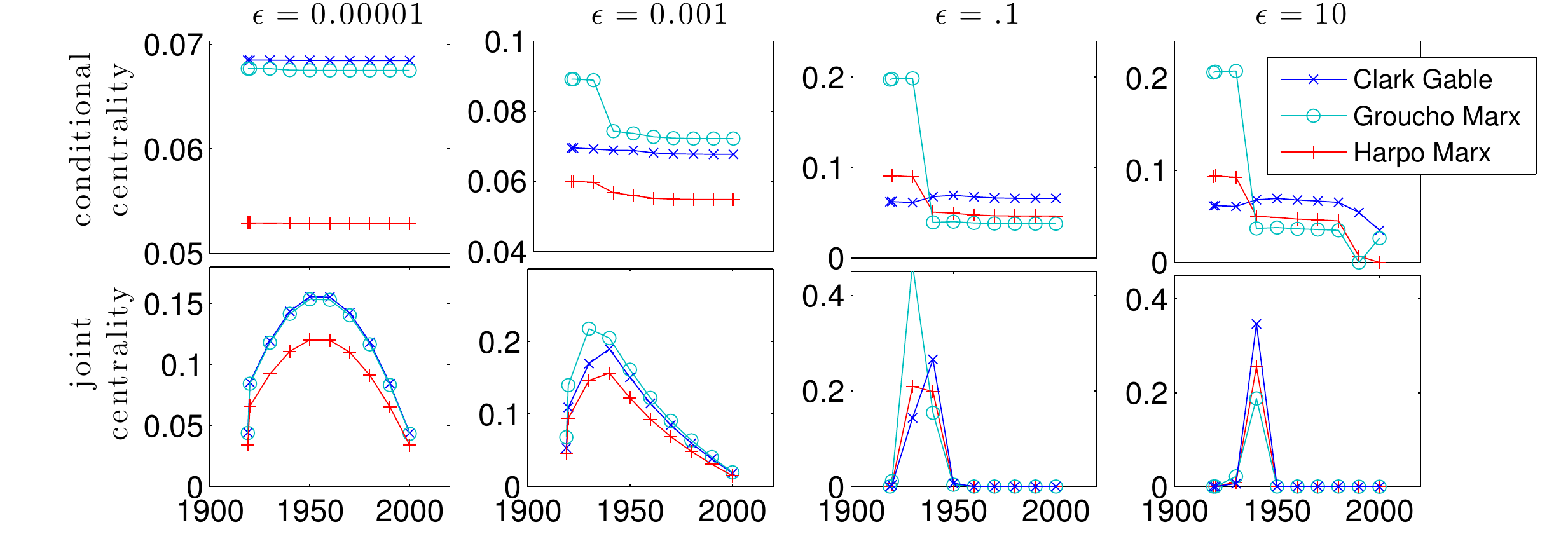}\\
\caption{
\drt{(Top) Conditional-centrality and (bottom) joint-centrality trajectories for the actors in Table \ref{table:GAH} with the highest time-averaged centralities. The columns depict centrality trajectories for various values of $\epsilon$, which increase from left to right. Note that although Clark Gable has the highest time-averaged centrality, there exist values of $\epsilon$ (e.g., $\epsilon=0.001$) for which Groucho Marx has higher conditional and joint centralities at all instances in time, indicating that there is insufficient evident in our study to conclude objectively which actor is more 
significant. 
}}
\label{fig:sensitive}
\end{figure}

\drt{Further study of the GAH network reveals additional insights about actor centralities. Specifically (see Table \ref{table:GAH}), Clark Gable has the highest time-averaged centrality $\alpha_i$ ($0.3683$), but the $\alpha_i$ value ($0.3627$) for Groucho Marx, and both of these values are much larger than that ($0.2844$) for the third-ranked actor, Harpo Marx. In Fig.~\ref{fig:sensitive}, we plot the conditional-centrality and joint-centrality trajectories for these three actors. The columns show centralities for several choices of $\epsilon$, which increase from left to right. For $\epsilon=0.00001$ (left column), the relative centralities---that is, the rank order of actors and the ratio of one actor's centrality versus any other's---are very similar to the limiting values (see Table \ref{table:GAH}) that one obtains for $\epsilon\to0$. However, there exist values of $\epsilon$ in which the centrality trajectory for Groucho Marx is consistently higher than that for Clark Gable. For example, see the second column (for which $\epsilon=0.001$) in Fig.~\ref{fig:sensitive}. This observation supports our belief that the examination of a single centrality measure (or, more generally, methodology for ranking) is insufficient to fully describe the importances of nodes in a network. 
In particular, ranking nodes based on any centrality calculation is sensitive to the choices that are made when defining the centrality.  Investigating the robustness of centrality measures is an important open question in network science (see e.g., \cite{Bloch}).
}

\begin{table}[t!]
\caption{Top time-averaged centralities \drt{($\alpha_i$)} and first-order-mover scores \drt{($m_i$)} for Supreme Court decisions \cite{SCD}. We give results for our temporal generalization [see Eq.~\eqref{eq:sup_centrality1}] of authority scores \cite{kleinberg1999}.}
\small{
\centering 
\begin{tabular}{c } 
Top Time-Averaged Centralities  \\
\begin{tabular}{c c c c}
\hline\hline 
Rank & Decision & $\alpha_i$ & $m_i$ \\ [0.5ex] 
\hline  
1 & Gibbons v. Ogden (22 U.S. 1, 1824) & 0.172 &\drt{284.8}  \\  
2 & Minnesota Rate Case (230 U.S. 352, 1913)  & 0.160&\drt{214.2}  \\
3 & McCulloch v. Maryland (17 U.S. 316, 1819) &  0.157 &\drt{173.3} \\
4 & Brown v. Maryland (25 U.S. 419, 1827) & 0.107 &\drt{212.7} \\
5 & Robbins v. Shelby County Tax. Distr. (120 U.S. 489, 1887) & 0.096  &\drt{192.0} \\ 
6 & Cooley v. Board of Wardens (53 U.S. 12, 1851)  & 0.096  &\drt{191.2} \\ 
7 & Ex parte Young (209 U.S. 123, 1908) & 0.086  &\drt{53.8} \\ 
8 & Galveston \& S.A. Ry. Co. v. Texas (210 U.S. 217, 1908) & 0.083 &\drt{126.6}  \\ 
9 & Cantwell v. Connecticut (310 U.S. 296, 1940) & 0.082 &\drt{516.2} \\ 
10 & Welton v. State of Missouri (91 U.S. 275, 1875) & 0.082  &\drt{168.5} \\ 
[1ex]  
\hline 
\end{tabular}
\\\\
Top First-Order-Mover Scores\\
\begin{tabular}{c c c c}  
\hline\hline  
Rank & Decision & $m_i$ & $\alpha_i$ \\ [0.5ex] 
\hline 
1 & Cantwell v. Connecticut (310 U.S. 296, 1940) & 516.21 &\drt{0.0822} \\ 
2 & Schneider v. State (308 U.S. 147, 1939) & 439.73 &\drt{0.0707} \\
3 & Thornhill v. Alabama (310 U.S. 88, 1940) & 388.08   &\drt{0.0633}\\
4 & Lovell v. City of Griffin (303 U.S. 444, 1938) & 369.00 &\drt{0.0632} \\
5 & Near v. Minnesota (283 U.S. 697, 1931) & 344.06  &\drt{0.0729} \\ 
6 & Gitlow v. New York (268 U.S. 652, 1925) & 316.33   &\drt{0.0626}\\ 
7 &  DeJonge v. Oregon (299 U.S. 353, 1937) & 310.63  &\drt{0.0554} \\ 
8 &  Stromberg v. California (283 U.S. 359, 1931) & 306.68   &\drt{0.0563}\\ 
9 &  Chaplinsky v. New Hampshire (315 U.S. 568, 1942) & 302.71 &\drt{0.0487} \\ 
10 & Whitney v. California (274 U.S. 357, 1927) & 291.03  &\drt{0.0582} \\ 
[1ex]  
\hline 
\end{tabular}
\end{tabular}
}
\label{table:SCD} 
\end{table}

\subsection{Citations of United States Supreme Court Decisions (SCD)}\label{sec:supremes}

In our final case study, we investigate the interconnectedness of Supreme Court decisions (SCD) in the Unites States by examining networks that encode citations of decisions \cite{leicht-citation2007,fowler2007b,fowler2008,bommarito2010}. Such an investigation can reveal a variety of insights about the decisions, including identifying which ones build on one another and illuminating the rise and fall of importance of decisions. One can also try to reveal insights into large-scale social processes during a given time period (e.g., the identification of which social issues are considered to be important and/or controversial).

We study the data set that was made available by Fowler et al.\drt{\cite{fowler_url,fowler2008}}, and we note that the complete decisions are available online from the U.S. government \cite{SCD}. We examine temporal citation networks for the time range 1800--2002, which we bin into decades to give $T=20$ time layers.\footnote{The final time layer is slightly longer than a decade, because it encompasses the years 1990--2002.} We construct a directed temporal network in which we include a directed edge from node $i$ to node $j$ at time $t$ if decision $i$ cites decision $j$ and decision $i$ was written during the $t$th decade. To study centrality in such a temporal network, we restrict our attention to the largest weakly-connected component, which contains $N=25,389$ nodes. We study our temporal generalization of authority scores \cite{kleinberg1999}, as high authority nodes should correspond to highly-cited, influential decisions.

In Table \ref{table:SCD}, we indicate the decisions that have the top time-averaged authority \drt{scores} and those that have the top first-order-mover authority scores. We identify the top-ranking decision to be Gibbons v. Ogden (22 U.S. 1, 1824), which is well-known to be a highly influential commerce case \cite{fowler2007b,fowler2008}. (Note that we use standard case notation, so for this example citation, 22 is the volume, 1 is the page, and 1824 is the year.) The decisions with the top time-average centralities tend to be decisions from before 1900. In contrast, the nodes with the top first-order-mover scores tend to be decisions from the period 1920--1940. These decisions initially have very low centralities because they do not exist early in the \drt{network}, but influential decisions from this period later achieve high authorities. For example, see Table 2 in Ref.~\cite{fowler2008}, which identified Cantwell v. Connecticut (310 U.S. 296, 1940) as the node with top authority for a network in which there exists an edge $i\to j$ if and only if decision $i$ cites decision $j$ during the years $t\in[1754,2002]$. This decision is the only one that makes both of our top-10 lists in Table \ref{table:SCD}.
\drt{Interestingly, the cases with top time-averaged centralities deal mostly with interstate commerce and the limits of state governments in regulating it, and the cases with top first-order-mover scores deal primarily with First-Amendment rights.}

\section{Conclusions}\label{sec:conclusion}

We developed and analyzed a generalization of eigenvector-based centrality measures for temporal networks, and we demonstrated the utility of such temporal centralities for identifying important entities in three case studies: the \drt{Ph.D. exchange network} for mathematical sciences in the United States, costarring in the Golden Age of Hollywood, and citations of decisions in the United States Supreme Court. Consistent with the lessons from the development of multilayer generalizations of \drt{modularity \cite{mucha2010,bassett2013,bazzi2015}, an} essential ingredient of generalizing centrality measures to multilayer representations of temporal networks is to give different treatment to intra-layer edges between different \drt{physical} nodes and inter-layer edges that connect the same node across time. We incorporated inter-layer edges by constructing a supra-centrality matrix, which requires a network to either have discrete time or be binned \drt{(i.e, aggregated)} into discrete times using time windows \drt{(e.g., using summation or another process \cite{taylor2016enhanced,taylor2016_b}).}
Constructing this matrix requires \drt{making a choice of centrality matrix (e.g., authority, hub, adjacency, PageRank, and so on), and one then couples these matrices} using an inter-layer coupling \drt{of} strength $\omega$ (i.e., $1/\epsilon$).

We showed that the dominant eigenvector of a supra-centrality matrix characterizes the \drt{``joint centralities''} of node-layer pairs, and we introduced the concepts of \drt{``marginal centralities''} and ``conditional centralities'' to study \drt{(1)} the decoupled importances of nodes and layers and \drt{(2) the nodes' centrality trajectories over time.} \drt{One needs to be cautious about interpreting joint centralities, as there is a bias towards the middle layers [e.g., see Figs.~\ref{fig:toy1}(a) and \ref{fig:GT1}(b)], but conditional centralities do not seem to have a similar bias. In our temporal centralities, we observe properties such as localization (which is an important issue for centrality measures more generally \cite{pastor2015distinct,kawamoto2015localized,hashimoto2015,taylor2016_b}) and different time scales for how node centrality changes over time.} For example, we observed eigenvector localization, amounting to localization in time layers, in the limit $\omega\to0$. Further research is important to explore eigenvector localization in temporal and multilayer networks and to examine the results from generalizing different types of eigenvector-based centralities.
It would also be worthwhile to explore whether concepts from statistics about smoothing, such as using cross-validation to choose bandwidth parameters, can be used to \drt{help} guide the selection of values of $\omega$.

By focusing on the strong-coupling limit---including the construction of a perturbation expansion in this singular limit---we derived simple, principled formulas to define time-averaged \drt{centralities and first-order-mover} scores (which measure the magnitude \drt{of} a physical node's centrality changes over time). This makes it possible to easily identify not only which entities are most central in a temporal network but also which ones are the ``top movers'' in centrality over time. Our methodology works for any eigenvector-based centrality, which we define as centrality measures in which the nodes' centralities are given by the entries of the dominant eigenvector of a matrix. There are numerous popular types of eigenvector-based centralities (including PageRank centralities \cite{pagerank}, hub and authority centralities \cite{kleinberg1999}, and eigenvector centrality \cite{bonacich1972}), and new centralities of this form continue to be developed \cite{hashimoto2015}.
\drt{We have also provided {\sc Matlab} software \cite{code} that implements our methodology. It can be used to extend any eigenvector-based centrality to temporal networks.}
 
An important direction \drt{for future work} is to explore different strategies for the construction of inter-layer edges. We have studied inter-layer edges between nearest-neighbor node-layer pairs; that is, a node-layer pair $(i,t)$ is adjacent to $(i,t-1)$ and $(i,t+1)$, so the edges form an undirected chain that bridges physical node $i$ across the $T$ time layers. Other strategies should be explored (see discussions in \cite{kivela2014}), including ones with inter-layer edges that are directed (e.g., other temporal generalizations of communicability centrality \cite{fenu2015}).  \drt{Directed inter-layer edges are able to represent causal scenarios, but such causal coupling} can yield \drt{(supra-centrality) matrices} that are not irreducible, which is a problematic situation for eigenvector-based centrality measures. Specifically, one would need to construct an associated network structure that satisfies strong connectivity (e.g., using ideas such as \drt{teleportation} \cite{gleich2014,lambiotte2012}). Developing eigenvector-based centrality measures that \drt{respect} causality is thus one \drt{exciting} direction, although one then needs to keep careful track of the biases caused by the choice of teleportation strategy. \drt{We note that we have developed our perturbation analyses for general inter-layer coupling, so they can be applied to such scenarios. Finally, it is worth exploring our approach for different applications such as change-point detection and for a variety of example networks such data streams and as those with a large number of time layer.}

\section*{Appendix: Higher-Order Terms}\label{sec:higher}

In Secs.~\ref{sec:perturb} and \ref{sec:movers}, we derived \drt{zeroth-order} and first-order solutions to the dominant-eigenvalue equation, Eq.~\eqref{eq:eval1}, for the supra-centrality matrix. These led to principled expressions for time-averaged centralities [see Eq.~\eqref{eq:ap1}] and first-order-mover scores [see Eq.~\eqref{eq:fmover}]. We now derive and solve higher-order terms in our singular perturbation expansion. Such terms can be useful for approximating $\mathbbm{v}(\epsilon)$ for fixed $\epsilon>0$. Similar to the expressions that we derived in Sec.~\ref{sec:TimeAve}, we obtain expressions in the form of linear equations of dimension $N$. For longitudinal data, these equations are much more computationally efficient to solve than directly solving Eq.~\eqref{eq:eval1}, which is an eigenvalue equation of dimension $NT$.

From the eigenvalue equation given by Eq.~\eqref{eq:eval1}, we now \drt{derive} $k$th-order expansions of the form $\lambda(\epsilon)\approx \sum_{j=0}^k \epsilon^k \lambda_k$ and $\mathbbm{v}(\epsilon)\approx \sum_{j=0}^k \epsilon^k \mathbbm{v}_k$. We can consider \drt{arbitrarily} large nonnegative integers $k$. We derived zeroth-order ($k = 0$) and first-order ($k = 1$) approximations in Sec.~\ref{sec:TimeAve}, and we now derive the second-order ($k=2$) expansion. Because we already showed that $\lambda_2 =  \bm{\alpha}^T \mathbf{X}^{(2)} \bm{\alpha}$ in Eq.~\eqref{eq:lam2}, all that is left is to derive an expression for $\mathbbm{v}_2$. Starting from Eq.~\eqref{eq:second0}, we write 
\begin{equation}
	\mathbbm{v}_2 = \mathbb{L}_0^\dagger\mathbb{L}_1\mathbbm{v}_1 + \sum_j \gamma_j \mathbb{P}\mathbbm{u}_j \,, \label{eq:second1}
\end{equation}
where the sum with constants $\gamma_j =(\mathbb{P}\mathbbm{u}_j)^T \mathbbm{v}_2$ accounts for the projection of $\mathbbm{v}_2$ onto the null space of $\mathbb{L}_0^\dagger$. We solve for the constants $\{\gamma_j\}$ by examing the third-order ($k = 3$) expansion, which leads to 
\begin{equation}
	\mathbb{L}_0\mathbbm{v}_3 = \mathbb{L}_1\mathbbm{v}_2 - \lambda_2 \mathbbm{v}_1 - \lambda_3 \mathbbm{v}_0 \,. \label{eq:third0}
\end{equation}
In general, 
\begin{equation}
	\mathbb{L}_0\mathbbm{v}_k = \mathbb{L}_1\mathbbm{v}_{k-1} - \sum_{j=2}^k \lambda_j \mathbbm{v}_{k-j} \,. \label{eq:kth0}
\end{equation}
We left-multiply Eq.~\eqref{eq:third0} by $(\mathbb{P}\mathbbm{u}_i)^T$ and note that $\mathbb{P}\mathbbm{u}_i$ is in the null space of $\mathbb{L}_0$ to obtain
\begin{align}
	0 &=(\mathbb{P}\mathbbm{u}_i)^T\mathbb{L}_1\mathbbm{v}_2  - \lambda_2 \beta_i - \lambda_3 \alpha_i\,. \label{eq:third1}
\end{align}
Using the solution for $\mathbbm{v}_2$ given by Eq.~\eqref{eq:second1}, it then follows that
\begin{align}
	-\sum_j \gamma_j (\mathbb{P}\mathbbm{u}_i)^T\mathbb{L}_1\mathbb{P}\mathbbm{u}_j   &=  (\mathbb{P}\mathbbm{u}_i)^T\mathbb{L}_1\mathbb{L}_0^\dagger\mathbb{L}_1\mathbbm{v}_1 - \lambda_2 \beta_i -\lambda_3 \alpha_i   \,.\label{eq:third2}
\end{align}
Recalling that $(\mathbb{P}\mathbbm{u}_i)^T\mathbb{L}_1\mathbb{P}\mathbbm{u}_j = {X}^{(1)}_{ij} -\lambda_1\delta_{ij}$, Eq.~\eqref{eq:third2} becomes 
\begin{align}
	-\left(\mathbf{X}^{(1)} - \lambda_1\mathbf{I}\right)\bm{\gamma}  &=  \bm{q}  - \lambda_2 \bm{\beta} - \lambda_3 \bm{\alpha} \,,\label{eq:third3}
\end{align}
where $\bm{q} = [q_1,\dots,q_N]^T$ and $q_i = (\mathbb{P}\mathbbm{u}_i)^T\mathbb{L}_1\mathbb{L}_0^\dagger\mathbb{L}_1\mathbbm{v}_1$. We now solve for $\lambda_3$. Using the fact that $\bm{\alpha}$ is in the null space of $\left(\mathbf{X}^{(1)} - \lambda_1\mathbf{I}\right)$, left-multiplication of Eq.~\eqref{eq:third3} by $\bm{\alpha}^T$ gives  
\begin{align}
	\lambda_3 &=  \bm{\alpha}^T\mathbf{q} = \mathbbm{v}_0^T \mathbb{L}_1 \mathbb{L}_0^\dagger \mathbb{L}_1\mathbbm{v}_1\,,\label{eq:third4}
\end{align}
where we have used the relation $\bm{\alpha}^T\bm{\beta}=0$.
As before, we solve for $\bm{\gamma}$ using the Moore--Penrose pseudoinverse:
\begin{align}
	\bm{\gamma}  &=  -\left(\mathbf{X}^{(1)} - \lambda_1\mathbf{I}\right)^\dagger \left(\mathbf{q}  - \lambda_2  \bm{\beta} - \lambda_3 \bm{\alpha}  \right) + c\bm{\alpha}\,,
	\label{eq:third5}
\end{align}
where $c=\bm{\alpha}^T\bm{\gamma}$ is the (possibly nonzero) projection of $\bm{\gamma}$ onto $\bm{\alpha}$. 

Similar to our calculation of $b=\bm{\alpha}^T\bm{\beta}=0$ in Sec.~\ref{sec:movers}, we solve for the constant $c$ by examining the norm of the vector $\mathbbm{v}(\epsilon) \approx \mathbbm{v}_0 + \epsilon \mathbbm{v}_1 + \epsilon^2 \mathbbm{v}_2$. We require that
\begin{align}
	1 &= \| \mathbbm{v}_0 + \epsilon \mathbbm{v}_1 +  \epsilon^2 \mathbbm{v}_2 \|^2\nonumber\\
&= \| \mathbbm{v}_0\|^2 +  \epsilon^2 \|\mathbbm{v}_1 \|^2
+  2\epsilon \langle\mathbbm{v}_0,\mathbbm{v}_1\rangle + 2\epsilon^2 \langle\mathbbm{v}_0,\mathbbm{v}_2\rangle + O(\epsilon^3) 
\,.\label{eq:norm3}
\end{align}
We substitute $\| \mathbbm{v}_0\|=1$ and $\langle\mathbbm{v}_0,\mathbbm{v}_1\rangle  = \bm{\alpha}^T\bm{\beta}=0$ into Eq.~\eqref{eq:norm3} and equate the $\mathcal{O}(\epsilon^2)$ terms to obtain
\begin{align}
	c &= \langle\mathbbm{v}_0,\mathbbm{v}_2\rangle =   -\frac{1}{2}\|\mathbbm{v}_1 \|^2 \, .\label{eq:norm4}
\end{align}
This completes the second-order approximation. 

We now derive the third-order ($k=3$) approximation. Having solved $\lambda_3$ in Eq.~\eqref{eq:third4}, we seek to find an expression for $\mathbbm{v}_3$. Starting from Eq.~\eqref{eq:third0}, we write 
\begin{equation}
	\mathbbm{v}_3 = \mathbb{L}_0^\dagger\mathbb{L}_1\mathbbm{v}_2 - \lambda_2\mathbb{L}_0^\dagger\mathbbm{v}_1 + \sum_j \xi_j \mathbb{P}\mathbbm{u}_j \,, \label{eq:third6}
\end{equation}
where the sum with constants $\xi_j = (\mathbb{P}\mathbbm{u}_j)^T \mathbbm{v}_3$ accounts for the projection of $\mathbbm{v}_3$ onto the null space of $\mathbb{L}_0^\dagger$. Using the general form given by Eq.~\eqref{eq:kth0} for $k=4$ yields 
\begin{equation}
	\mathbb{L}_0\mathbbm{v}_4 = \mathbb{L}_1\mathbbm{v}_3 - \lambda_2 \mathbbm{v}_2 - \lambda_3 \mathbbm{v}_1 - \lambda_4 \mathbbm{v}_0\label{eq:fourth0} \,.
\end{equation}
We left-multiply Eq.~\eqref{eq:fourth0} by $(\mathbb{P}\mathbbm{u}_i)^T$ to obtain
\begin{equation}
	0 = (\mathbb{P}\mathbbm{u}_i)^T\mathbb{L}_1\mathbbm{v}_3 - \lambda_2 \gamma_i - \lambda_3 \beta_i - \lambda_4 \alpha_i \,.
	\label{eq:fourth1} 
\end{equation}
We next substitute our solution for $\mathbbm{v}_3$ given by Eq.~\eqref{eq:third6} into Eq.~\eqref{eq:fourth1} to obtain
\begin{equation}
	- \sum_j \xi_j (\mathbb{P}\mathbbm{u}_i)^T\mathbb{L}_1 \mathbb{P}\mathbbm{u}_j= 
(\mathbb{P}\mathbbm{u}_i)^T\mathbb{L}_1\mathbb{L}_0^\dagger \left(\mathbb{L}_1\mathbbm{v}_2 - \lambda_2\mathbbm{v}_1 \right)
	 - \lambda_2 \gamma_i - \lambda_3 \beta_i - \lambda_4 \alpha_i \label{eq:fourth2} \,,
\end{equation}
which in matrix notation is
\begin{equation}
	(\mathbf{X}^{(1)} - \lambda_1\mathbf{I})\bm{\xi} = \mathbf{r} - \lambda_2 \bm{\gamma} - \lambda_3 \bm{\beta} - \lambda_4 \bm{\alpha} \label{eq:fourth3} \,,
\end{equation}
where we define $\mathbf{r} = [r_1,\dots,r_N]^T$ and $r_i = (\mathbb{P}\mathbbm{u}_i)^T\mathbb{L}_1\mathbb{L}_0^\dagger \left(\mathbb{L}_1\mathbbm{v}_2 - \lambda_2\mathbbm{v}_1 \right)$.
As before, we use the Moore--Penrose psuedoinverse to obtain 
\begin{equation}
	\bm{\xi} = (\mathbf{X}^{(1)} - \lambda_1\mathbf{I})^\dagger \mathbf{r} - \lambda_2 (\mathbf{X}^{(1)} - \lambda_1\mathbf{I})^\dagger \bm{\gamma} - \lambda_3 (\mathbf{X}^{(1)} - \lambda_1\mathbf{I})^\dagger \bm{\beta}  + d\bm{\alpha}\,, 
		\label{eq:fourth4}
\end{equation}
which uses the fact that $\bm{\alpha}$ is in the null space of $(\mathbf{X}^{(1)} - \lambda_1\mathbf{I})$. The constant $d = \bm{\alpha}^T\bm{\xi}$ gives the projection of $\bm{\xi}$ onto $\bm{\alpha}$, and we calculate it by constraining the third-order expansion for $\mathbbm{v}(\epsilon)$ to have a norm of $1$. This yields
\begin{align}
	1 &= \| \mathbbm{v}_0 + \epsilon \mathbbm{v}_1 +  \epsilon^2 \mathbbm{v}_2 + \epsilon^3 \mathbbm{v}_3 \|^2\nonumber\\
	&= \| \mathbbm{v}_0\|^2 +  \epsilon^2 \|\mathbbm{v}_1 \|^2 
+ 2\epsilon \langle\mathbbm{v}_0,\mathbbm{v}_1\rangle + 2\epsilon^2 \langle\mathbbm{v}_0,\mathbbm{v}_2\rangle  \nonumber\\
	&~ + 2\epsilon^3 \langle\mathbbm{v}_0,\mathbbm{v}_3\rangle + 2\epsilon^3 \langle\mathbbm{v}_1,\mathbbm{v}_2\rangle 
+ O(\epsilon^4) \,.
	\label{eq:norm6}
\end{align}
Recall that normalization of the zeroth-, first-, and second-order expansions, respectively, \drt{yields} the relations $\|\mathbbm{v}_0\|^2=1$, $\langle\mathbbm{v}_0,\mathbbm{v}_1\rangle=0$, and $\langle\mathbbm{v}_0,\mathbbm{v}_2\rangle = -(1/2)\|\mathbbm{v}_1\|^2$. Equating the third-order terms in Eq.~\eqref{eq:norm6} then necessitates that
\begin{align}
	d &= \langle\mathbbm{v}_0,\mathbbm{v}_3\rangle  =  - \langle\mathbbm{v}_1,\mathbbm{v}_2\rangle \,.
	\label{eq:norm7}
\end{align}
This completes the third-order approximation.

\begin{figure}[t!]
\centering
\includegraphics[width=\linewidth]{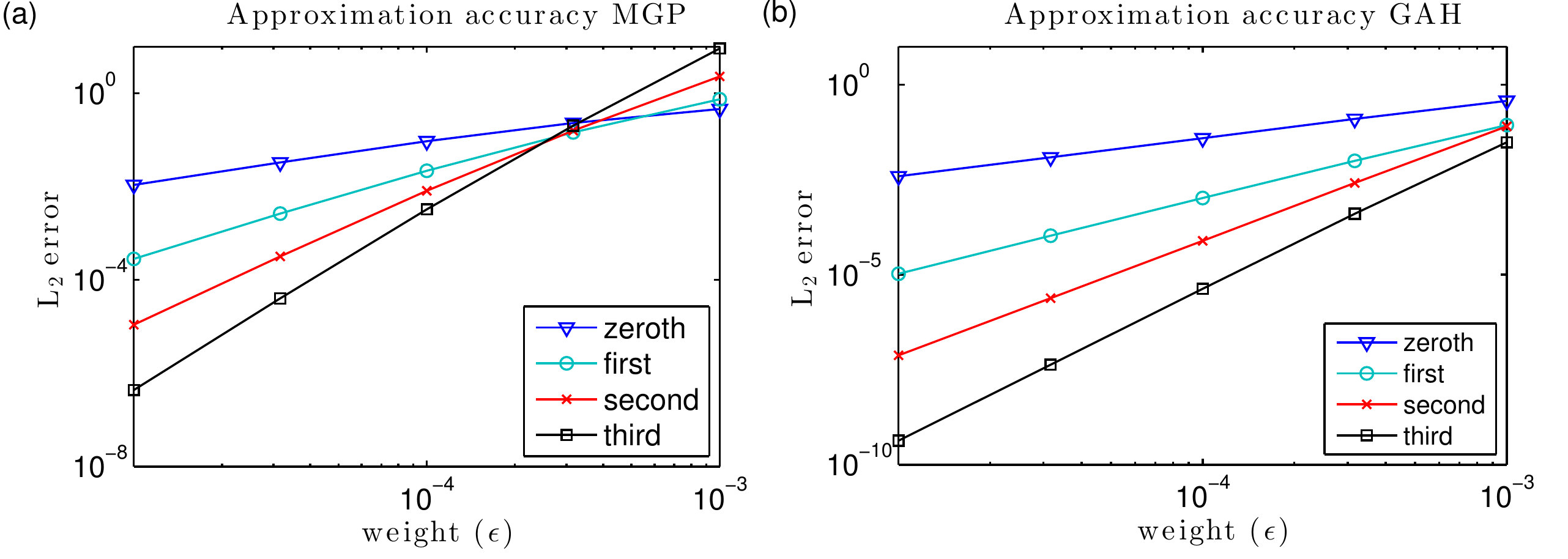}\\
\caption{We show, for several values of inter-layer coupling weight $\epsilon>0$, the accuracy of $k$th-order approximate solutions to Eq.~\eqref{eq:eval1} for $k\in\{0,1,2,3\}$. We show results for (a) the MGP network studied in Sec.~\ref{sec:mgp} and (b) the GAH network studied in Sec.~\ref{sec:hollywood}. In both panels, we measure the error by calculating the $L_2$ norm $\|\mathbbm{v}(\epsilon) -\sum_{j=1}^k \epsilon^j \mathbbm{v}_k\|_2$ of the difference between the approximate and actual dominant eigenvector. 
As expected, we find for sufficiently small $\epsilon$ that the error decreases with increasing approximation order. The decay rate of the $L_2$ error as $\epsilon\to0^+$ also follows the expected scaling.
}
\label{fig:higher}
\end{figure}

In Figs.~\ref{fig:higher}(a) and \ref{fig:higher}(b), we validate the above results using the MGP and GAH networks, which we studied in Sec.~\ref{sec:mgp} and Sec.~\ref{sec:hollywood}, respectively. We plot the $L_2$ error for the $k$th-order approximation $\|\mathbbm{v}(\epsilon) -\sum_{j=1}^k \epsilon^j \mathbbm{v}_k\|_2$ to the dominant eigenvector. We plot these approximations for several choices of $k$ and several values of $\epsilon$. When $\epsilon$ is sufficiently small (i.e., $\epsilon \lessapprox 3 \times 10^{-4}$ for the MGP and $\epsilon \lessapprox 10^{-3}$ for the GAH), we observe (as expected) that the error decreases with increasing $k$. We derived our approximate expressions in the limit $\epsilon\to0^+$, so we only expect them to be accurate for sufficiently small $\epsilon$ (although asymptotic expressions do have a long history of often being accurate even in many situations in which there are no guarantees for such success \cite{hinch1991}). We also obtain the expected decay rates in the error as $\epsilon\to0^+$. We find linear decay for the zeroth-order approximation, quadratic decay for the first-order approximation, and so on.

\bibliographystyle{siam}
\bibliography{tempcent5}

\end{document}